\documentclass[aps,prd,twocolumn,groupedaddress,nofootinbib,showpacs,eqsecnum,useAms]{revtex4-1}

\usepackage{hyperref}
\usepackage{epsfig}
\usepackage{verbatim}
\usepackage{euscript,graphicx}
\usepackage{amsthm,dsfont,amsfonts,amsmath,amssymb}
\usepackage{euscript,color,fontenc,textcomp,relsize}
\usepackage{bm,url,float,cleveref}
\usepackage[caption=false]{subfig}

\expandafter\ifx\csname package@font\endcsname\relax\else
 \expandafter\expandafter
 \expandafter\usepackage
 \expandafter\expandafter
 \expandafter{\csname package@font\endcsname}%
\fi
\allowdisplaybreaks

\begin{document}

\date{\today}

\title{Frequency and time domain inspiral templates for 
comparable mass compact binaries in eccentric orbits}

\author{
Sashwat Tanay, Maria Haney, and
Achamveedu Gopakumar\footnote{gopu@tifr.res.in} }

\affiliation{
 Tata Institute of Fundamental Research, Mumbai 400005, Maharashtra, India}

\begin{abstract}
  Inspiraling compact binaries
with non-negligible orbital eccentricities are  
 plausible gravitational wave (GW) sources for the
upcoming network of GW observatories.
In this paper, we present two prescriptions to 
compute post-Newtonian (PN) accurate inspiral templates 
for such binaries.
First, we adapt and extend  
the post-circular scheme of 
 Yunes {\it et al.} 
[Phys. Rev. D 80, 084001 (2009)] to obtain a Fourier-domain 
inspiral approximant that 
incorporates the effects of 
PN-accurate orbital eccentricity evolution.
This results in a 
fully analytic frequency-domain 
inspiral waveform with Newtonian amplitude 
and 2PN order  Fourier phase while incorporating 
eccentricity effects up to sixth order at each PN order.
The importance of incorporating eccentricity evolution contributions to
the Fourier phase in a PN consistent manner is also demonstrated.
Second, we present an accurate and efficient prescription to 
incorporate orbital eccentricity into the quasi-circular 
time-domain {\texttt{TaylorT4}} approximant at 2PN order. 
New features include the use of 
rational functions in orbital eccentricity to  
implement the 1.5PN order tail contributions to the far-zone
fluxes.
 This leads to  closed form PN-accurate differential
equations for evolving eccentric orbits and 
the resulting time-domain approximant is accurate and 
efficient to handle initial orbital eccentricities $\leq 0.9$.
 Preliminary GW data analysis implications are 
probed using  match estimates.

\end{abstract}


\maketitle

\section{Introduction}

  Coalescing comparable mass compact binaries are the expected 
workhorse GW sources for the upcoming second generation kilometer 
sized laser interferometer systems \cite{LR_SS}.
%http://relativity.livingreviews.org/Articles/lrr-2009-2/fulltext.html
For example, 
a network of GW observatories,  
such as Advanced LIGO (aLIGO) \cite{aLIGO}, Virgo \cite{aVirgo}, and Kagra \cite{Kagra},
may be able to detect roughly
$5-50$ GW events/year, provided 
the beamed short Gamma-ray bursts (GRBs) are due to
the merger of stellar mass compact binaries \cite{CH12}.
%http://arxiv.org/pdf/1206.0703v1.pdf
The stellar mass compact binaries, containing neutron stars (NSs)
and black holes (BHs), are expected to shed their formation eccentricities due to the gravitational radiation reaction \cite{PM63,P64}.
% Refs 8 & 9 in Tunes
 This is why the Hulse-Taylor binary pulsar whose 
present orbital period and eccentricity are $ \sim 8$ hr and $ \sim 0.6$, respectively \cite{taylornobel},
% Taylor Jr, J.H., “Nobel Lecture: Binary pulsars and relativistic gravity”, Rev. Mod. Phys., 66, 711–719, (1994).
will have an orbital eccentricity $\sim 10^{-6}$ when 
GW driven inspiral brings its orbital frequency to few Hertz.
Therefore, isolated compact binaries are expected to be in 
 quasi-circular orbits when they spiral into the frequency windows
of terrestrial GW observatories. 
This makes  coalescing compact binaries in quasi-circular orbits the most promising GW
sources for these observatories.
Additionally, 
 there exist several quasi-circular inspiral waveform families
 to extract these weak GW signals from the noisy data streams   \cite{BIOPS}.

        Very recently,  Huerta {\it et al.} argued that 
aLIGO type observatories could detect roughly $0.1-10$ eccentric 
inspirals per year up to redshift $z \sim 0.2$ \cite{H14}. 
This study was influenced by a number of recent investigations that explored 
several plausible astrophysical mechanisms for 
producing aLIGO relevant compact binaries with non-negligible eccentricities. These observationally unconstrained 
short period compact binary formation
scenarios include  
dynamical capture in dense stellar environments, present in  both galactic central regions and globular 
clusters,  as well as tidal capturing of compact objects by neutron stars 
( see for example Refs.~\cite{Samsing14,OKL09,LRV10} ).
%Refs 8,11, 31 in H14
Detailed listing of various astrophysical 
scenarios and mechanisms for producing ultra compact eccentric 
binaries that will retain residual eccentricities for both ground 
and space-based GW observatories 
can be found in Refs.~\cite{H14,YABW09,KG06}.
% http://adsabs.harvard.edu/abs/2006PhRvD..73l4012K
Therefore, it is possible that 
 the data streams of aLIGO type observatories 
may contain  
GWs from eccentric inspirals. 
This motivated us to construct PN-accurate and 
computationally efficient waveform families 
to model inspiral GWs from compact binaries in eccentric 
orbits.

  In the case of non-spinning compact binaries inspiraling 
along quasi-circular orbits, 
 {\texttt{TaylorT4}} and {\texttt{TaylorF2}} are popular 
models that 
 provide appropriate interferometric response functions
in the time and frequency domain, respectively \cite{Boyle_T4,BIOPS}.
These models, usually termed as approximants \cite{DIS98}, use PN approximation to general relativity 
to describe the frequency and phase evolution of GWs from 
compact binaries \cite{LR_LB}.
The  PN approximation is also employed 
to specify the amplitudes of the two polarization
states, namely $h_{\times}$ and $h_+$.
% in these approximants 
In this context,
the PN approximation provides general relativity based 
corrections to the Newtonian (or quadrupolar) estimates
such that $n$PN corrections give contributions 
that are accurate up to the relative $(v/c)^{2n}$ order beyond their
 Newtonian estimate, where $v$ and $c$ are the orbital and light 
speeds, respectively.
To incorporate higher order  PN  contributions to the frequency 
evolution as well as to the amplitudes of $h_{\times}$ and $h_+$, these approximants 
employ the dimensionless post-Newtonian expansion 
parameter $x$ \cite{BDI}.
This gauge-invariant parameter is defined as
$x  \equiv { \left(   {G m \omega} / {c^3}  \right) }^{2/3}$,
where $m$ is the total binary mass while $\omega$ stands 
for the orbital (angular) frequency.
Currently,
the state of the art 3.5PN order {\texttt{TaylorT4}} approximant 
provides corrections to the frequency evolution that are 
 accurate to ${\cal O}(x^{3.5})$ beyond the quadrupolar (Newtonian)
estimate \cite{BDEI04}.
In contrast,
the fully analytic 3.5PN order {\texttt{TaylorF2}} 
 approximant employing 3.5PN-accurate
Fourier phase
 is widely used to model quasi-circular inspiral templates
in the frequency domain.
At present, the amplitude
corrected GW polarization states 
for compact binaries in circular orbits
are available to the relative 3PN order \cite{BFIS08}.
Unfortunately, quasi-circular inspiral waveforms are 
substantially suboptimal to detect GWs from compact 
binaries with residual eccentricity $> 0.05$ \cite{HB2013}. 
In light of the above discussions, 
it should be of definite interest to extend 
these approximants by including the effects of non-negligible 
orbital eccentricities.

  In the first part of the paper, we incorporate the effects of 
PN-accurate orbital eccentricity evolution into the Fourier
phase of the 2PN-accurate circular 
{\texttt{TaylorF2}} 
 approximant.
This is done by adapting and extending 
the post-circular (PC) scheme of  
 Yunes {\it et al.} \cite{YABW09}. 
 In this approach, one first computes eccentricity induced higher $\omega$-based
  harmonic corrections
to both the amplitudes and  phases 
of the two  
 GW polarization states.
This leads to  the data analysis
relevant {\it response function} $h(t)$
for compact binaries inspiraling along eccentric orbits.
The frequency-domain inspiral templates were constructed by invoking 
the {\it Stationary Phase Approximation (SPA)},
detailed in Ref.~\cite{BO99_book}, on this $h(t)$.
To obtain fully  analytic frequency-domain inspiral templates,
Ref.~\cite{YABW09} also adapted and extended the idea of 
an asymptotic eccentricity invariant, introduced in Ref.~\cite{KKS95}.
This allowed  Yunes {\it et al.} to write down an analytic 
expression for the orbital eccentricity in terms of 
$\omega$, its initial value $\omega_0$ 
and $e_0$, the value of orbital eccentricity at $\omega_0$,
such that the uncontrolled errors are of ${\cal O}(e_0^9)$.
With the help of these ingredients, 
Ref.~\cite{YABW09} explicitly computed $\tilde h(f)$,
the frequency-domain version of $h(t)$,
 while incorporating only the Newtonian (quadrupolar) order 
contributions both in the 
amplitudes and phases 
of the two  
 GW polarization states.
 Very recently, Ref.~\cite{H14} developed an 
enhanced post-circular (EPC) formalism
to extend 
the PN-accuracy of the Fourier phase of the above $\tilde h(f)$.
For this purpose, Huerta {\it et al.}
 computed certain 3.5PN order eccentric contributions to the Fourier phase of the circular 
 {\texttt{TaylorF2}} 
 approximant. 
This was done by defining an eccentricity 
dependent velocity function 
from the quadrupolar order Fourier phase of Ref.~\cite{YABW09}.
This modified velocity function was then incorporated into
 the Fourier phase expression for the 3.5PN-accurate quasi-circular {\texttt{TaylorF2}} 
approximant. Indeed,
it was noted in Ref.~\cite{H14} that the EPC model 
does not provide a
 consistent PN extension  of the Newtonian order Fourier phase of 
 the PC scheme.
 However, its simplicity 
turned out 
 to be very useful for pursuing preliminary GW data analysis and astrophysical implications
associated with detecting eccentric inspirals with aLIGO
type observatories.
Notably,
the 3.5PN order EPC $\tilde h(f)$ was employed 
 to show that aLIGO could observe $\sim 0.1-10$ eccentric inspirals per year 
 out to $z \sim 0.2$ \cite{H14}.
%{ \bf Additionally,  a line about MT papers}.

     In this paper, we provide an approach to 
incorporate eccentricity contributions 
in a PN consistent manner 
to the quadrupolar order $\tilde h(f)$ of Ref.~\cite{YABW09}.
This is mainly achieved by 
incorporating the effects of 
PN-accurate orbital eccentricity evolution into the Fourier
phase of the above $\tilde h(f)$.
A crucial ingredient is the derivation of a 2PN-accurate 
expression for a certain orbital eccentricity $e_t$ as a bivariate expansion in terms of 
$x$ and $e_0$.
The eccentricity parameter $e_t$, referred to as the time-eccentricity,
appears in the Keplerian-type parametric solution to the 
PN-accurate compact binary dynamics \cite{DD85,MGS}.
This parameter is required to characterize the orbital
ellipticity while modeling GWs from compact binaries
inspiraling along PN-accurate eccentric orbits \cite{DGI}.
With the help of our 2PN-accurate expression for 
$e_t$ and Ref.~\cite{YABW09}, we derive the PN-accurate 
Fourier phase of $\tilde{h}(f)$.
 The resulting fully analytic frequency-domain approximant provides 
inspiral waveforms with Newtonian amplitudes 
and 2PN order Fourier phase while incorporating 
eccentricity evolution contributions accurate
 up to sixth order in $e_0$ at each PN level.
 
 To check the accuracy of our approach, we first explore how 
our approximate analytic $e_t$ estimates differ from their 
2PN-accurate numerical $e_t(\omega)$ counterparts that 
treat $e_t$ in an exact manner.
The maximum differences turned out to be  $ \leq 2\%$ of their 
initial values $e_0$
for a wide variety of binary and orbital parameters even during the late inspiral. 
We also show the importance of incorporating eccentricity 
contributions to the Fourier phase in a PN consistent manner.
This is essentially achieved by computing three different estimates 
for the accumulated GW cycles ($\mathcal{N}$)
in the aLIGO frequency window while 
using the $l=2$ harmonics of
 eccentric inspirals. 
 In this paper, the term `aLIGO frequency window' is
 used to indicate the  
 lower and upper limits for $x$, namely
 $x_{\rm low} ={ \left(   {G\, m\, \pi\, 10 } / {c^3}  \right) }^{2/3}$
 and $x_{\rm high}= 1/6$.
This indicates that we let the orbital evolution begin from
a fiducial GW frequency of $10$ Hz and end it at the last stable 
orbit of the binary, specified by $6 \, G\, m/c^2$.
Let us stress that for brevity we henceforth 
use `aLIGO frequency window'  as a short hand to 
denote the limits of binary evolution in the frequency windows of various advanced GW 
observatories like aLIGO, Virgo and Kagra.
The above mentioned
 $\mathcal{N}$ estimates arise from three different 
analytic expressions for the orbital phase $\phi (\omega, \omega_0,e_0)$ as well as
 our eccentric extension of the 2PN-accurate {\texttt{TaylorT4}} approximant, detailed 
in Sec.~\ref{sec:eT4}, that treats $e_t$ effects in an exact manner.
The analytic expressions for 
$\phi (\omega, \omega_0,e_0)$ are based on our 2PN-accurate expression 
for $e_t$, the EPC approach and the PC approach, supplemented
by the 2PN-accurate circular expression for $\phi (\omega)$.
We find that 
the $\mathcal{N}$ estimates, based on our PN-accurate
$\phi (\omega, \omega_0,e_0)$,
 are comparatively closer to those estimates arising from 
the eccentric extension of the 2PN-accurate {\texttt{TaylorT4}} approximant.
This is a desirable feature, as our time-domain eccentric approximant 
can be treated as an improved version of the $x$-model 
which was calibrated against a numerical relativity
simulation in Ref.~\cite{Hinder09}.  
In our view, this also points to the importance of including 
eccentricity contributions in a PN consistent manner while computing Fourier-domain inspiral templates for eccentric inspirals.
However, it will be desirable to include explicitly
PN effects due to periastron advance, higher order radiation reaction 
and spins into our analytic $\tilde h(f)$.
The resulting $\tilde h(f)$ should be useful to construct
computationally efficient PN-accurate Fourier-domain 
search templates for compact binaries in inspiraling 
eccentric orbits.

In the second part of the paper, we 
 describe our prescription to include the effects of 
 orbital eccentricity into the time-domain {\texttt{TaylorT4}} approximant
 in an accurate and efficient manner.
This approximant turned out to be the natural candidate 
for incorporating eccentricity effects in an efficient and 
exact manner among various time domain circular approximants like 
{\texttt{TaylorT1}}, {\texttt{TaylorT2}}, {\texttt{TaylorT3}} 
and {\texttt{TaylorT4}}.
We adapt the phasing formalism, detailed in Ref.~\cite{DGI},
 while employing the gauge-invariant $x$ parameter as
 done in Ref.~\cite{Hinder09}.
   This approach systematically incorporates the fact that 
GW signals emitted by compact binaries in  
inspiraling eccentric
orbits contain {\it three different time scales}, namely the orbital,
periastron precession and radiation-reaction time scales.
In the present implementation, the orbital dynamics 
is fully 2PN-accurate. Therefore, this time-domain approximant 
models GWs from non-spinning compact binaries 
that move along 2PN-accurate precessing eccentric orbits 
while inspiraling under the influence of GW emission that 
is fully 2PN 
accurate.
We provide four PN-accurate differential equations to
incorporate  secular variations, both conservative and 
dissipative, into the orbital variables that are present 
in the PN-accurate expressions for the two GW polarization states.
In contrast, 
the orbital time scale variations are included 
with the help of the 2PN-accurate Keplerian-type parametric solution in harmonic gauge \cite{MGS}.
The use of a modified Mikkola's method, detailed in Ref.~\cite{SM87}, to solve the 2PN-accurate Kepler 
equation ensures that orbital time scale variations 
in the dynamical variables are 
implemented in an accurate and computationally inexpensive way.
Another new feature is the use of {\it rational functions}
in orbital eccentricity
to incorporate the leading order tail contributions to the dissipative 
dynamics.
 This allows us to replace the infinite sum of Bessel functions
in terms of which the 1.5PN order tail contributions to the far-zone fluxes are usually specified \cite{BS93,RS97}.
The use of rational functions ensures that our approach 
can tackle initial eccentricities $\leq 0.9$ in a 
computationally efficient way.
We briefly contrast our approach with the
the  $x$-model of Ref.~\cite{Hinder09} and point out that further investigations 
will be required to estimate  the comparative accuracies  and efficiencies of the two approaches.
With the help of match estimates, we show that our time-domain prescription that treats eccentricity 
in an exact manner should be required to faithfully capture eccentric inspirals 
with $e_0 \geq 0.2$.

The paper is organized in the following way. In Sec.~\ref{sec:eTF2} we 
present our approach to incorporate PN order 
eccentric contributions to
the Fourier phase of $\tilde h(f)$ given in Ref.~\cite{YABW09}
and probe its salient features.
The formalism with which we incorporate the effects of orbital eccentricity
into the time-domain quasi-circular {\texttt{TaylorT4}} approximant
is detailed in Sec.~\ref{sec:eT4}.
We also probe preliminary data analysis implications of 
our approximant in this section.
A brief summary, possible implications and extensions are listed in Sec.~\ref{sec:conclusions}.

\section{Analytic $\tilde h(f)$ for eccentric inspirals with 2PN order Fourier phase }\label{sec:eTF2}

 We begin with a brief review of the PC approach of Ref.~\cite{YABW09}
  to compute an analytic frequency-domain response function 
 with quadrupolar (Newtonian) order amplitude and phase
 for eccentric inspirals.
The extension of this approach  to obtain analytical $\tilde h(f)$ with 2PN-accurate 
Fourier phase and its preliminary implications are 
 presented in Sec.~\ref{sectionB}.
In this extension, we focus on the effect of PN-accurate 
eccentricity (and frequency) evolution on the  Fourier phase.
However, the influence of periastron advance on the harmonic structure 
of GW polarizations states, as explored in Ref.~\cite{TG07}, 
and its influence on  $\tilde h(f)$
is not pursued in the present extension.

\subsection{Newtonian order post-circular $\tilde h(f)$ }
\label{sec:eTF2_1}

The approach of 
Ref.~\cite{YABW09} begins by 
expressing  $h_{\times}$ and $h_{+}$ 
for compact binaries in eccentric orbits 
as a sum over harmonics. These harmonics are defined in terms 
of the mean anomaly $l= 2\, \pi\, F \, (t -t_0)$, where 
$F$ is the orbital frequency while $t_0$ is some initial epoch.
The quadrupolar (Newtonian) order expressions for the 
two polarization states that incorporate eccentricity 
corrections up to ${\cal O}(e_t^8)$, given in Ref.~\cite{YABW09},  take the form 
\begin{widetext}
\begin{equation}       
h_{+,\times} (t)=    -   \frac{G m \eta}{c^2 D_L}   \, x\,
%\left(\frac{G m 2 \pi F}{c^3} \right)^{2/3}    
   \sum\limits_{j=1}^{10}    \left[   C_{+,\times}^{(j)} \cos{ j l}  + S_{+,\times}^{(j)} \sin{j l}    \right]\,,     \label{1}        
\end{equation}
\end{widetext}
where $\eta$ and $D_L$ stand for the symmetric mass ratio and the luminosity distance, respectively.
The symmetric mass ratio $\eta$ of a binary consisting of individual masses 
$m_1$ and $m_2$ is defined to be $\eta= (m_1\,m_2)/ m^2$, where the total mass  
$m= m_1 + m_2$.
% Also $G$ and $c$ stand for the gravitational constant and the speed of light respectively.
The amplitudes 
 $C^{(j)}_{\times,+}$ and $S^{(j)}_{\times,+}$ are 
power series in $e_t$ whose coefficients are  
  trigonometric functions of the two angles 
$\iota, \beta$ that specify the line of sight vector in a certain
inertial frame.
Recall that the time eccentricity parameter $e_t$ is identical 
to the usual orbital eccentricity at the Newtonian order.
% associated with the initial (fixed) direction of 
%the orbital angular momentum of the binary.
The explicit expressions for these amplitudes, accurate up to
${\cal O}(e_t^8)$, are provided by Eqs.~(B) in Ref.~\cite{YABW09}.
The above two expressions for $h_{\times,+}$ arise from the Newtonian order GW polarizations, derived in Ref.~\cite{HW1987},
in terms of the orbital eccentricity and trigonometric functions 
of the true anomaly $\phi, \iota$ and $\beta$.
The harmonic structure of Eq.~(\ref{1}) is obtained with the help of 
infinite series expansions for $\sin \phi$ and $\cos \phi$ 
in terms of  $\sin j\,l$ and $\cos j\,l$.
The coefficients of $\sin j\,l$ and $\cos j\,l$ involve
orbital eccentricity and the Bessel functions of first kind  $J_j( j\, e_t)$.
The explicit harmonic content of Eq.~(\ref{1}) is the result of Taylor expanding the
eccentricity factors and 
$J_j( j\, e_t)$ in the small eccentricity limit.

  The detector strain or interferometric response function for GWs is 
  defined to be 
\begin{widetext}
\begin{equation}
h(t)= F_+\left(\theta_S,\phi_S,\psi_S\right)  h_+(t) +  F_\times\left(\theta_S,\phi_S,\psi_S\right) h_\times (t) \,,  \label{2}  
\end{equation}
\end{widetext}
where $F_{\times,+}\left(\theta_S,\phi_S,\psi_S\right)$ 
are the two detector antenna patterns.
These quantities depend on the right ascension
and declination of the source as well as the polarization angle 
$\psi_S$ \cite{KT89_300}.
With the above equations for $h_{\times}$ and $h_{+}$, it is fairly 
 straightforward to obtain 
$h(t)$ for GWs from compact binaries in eccentric orbits.
The resulting expression for the response function
for eccentric inspirals, given by Eq.~(4.21) of Ref.~\cite{YABW09}, reads
\begin{equation}
h(t) =    -   \frac{G m \eta}{c^2 D_L}   \left(\frac{G m \omega}{c^3} \right)^{2/3}   \sum\limits_{j=1}^{10}  \alpha_j~\cos\left(j l +\phi_j\right) ,     \label{3}     
\end{equation} 
where $\alpha_j =  {\rm sgn}(\Gamma_j) \sqrt{\Gamma_j^{2}+\Sigma_j^{2}}  $ and $ \phi_j = \arctan{\left(-\frac{\Sigma_j}{\Gamma
_j}\right)} $. 
The two new functions, 
$ \Gamma_j$ and $ \Sigma_j$,  are defined as
$\Gamma_j=
 F_+\, C^{(j)}_+ + F_{\times}\, C^{(j)}_{\times}$
and $\Sigma_j =   F_+\, S^{(j)}_+ + F_{\times}\, S^{(j)}_{\times}  $, respectively. 
Note that the above $\phi_j$
should not be confused with  the true anomaly $\phi$.
To obtain $h(t)$ for GWs from inspiraling 
binaries in 
eccentric orbits, we 
need to specify how $\omega = 2\, \pi\, F$ and $e_t$ vary in time.
At the  quadrupolar order, the temporal evolution 
for $\omega$ and $e_t$ is defined by 
%This is achieved by numerically solving the following  system of two %coupled differential equations:
% which at the quadrupolar order, available in Ref~\cite{}, read
%
\begin{widetext}
\begin{subequations}         \label{4}
\begin{align}           
\frac{d \omega}{dt} &=
\frac { (G\,m\ \omega)^{5/3}\,\omega^{2}\,\eta}
{5\,c^5\, (1 -e_t^2)^{7/2}}
\biggl \{  96
+292\,{{e_t}}^{2}
+37\,{{e_t}}^{4} \biggr \}          ,           \label{4.b}\\    
\frac{d e_t}{dt} &=
-\frac { (G\,m\, \omega)^{5/3}\, \omega \,\eta \, e_t }
{ 15\,c^5\, (1 -e_t^2)^{5/2}}
\biggl \{ 304 +  121\,{{e_t}}^{2} \biggr \} \,,        \label{4.c}
\end{align}
\end{subequations}
\end{widetext}
%{\bf See Eqs at the end of the document}
 adapted from Refs.~\cite{PM63,JS92}.
Clearly, we need to solve these two coupled differential equations 
numerically to obtain $\omega(t)$ and $e_t(t)$. This makes the procedure to obtain $h(t)$ for GWs from inspiraling eccentric 
binaries computationally expensive compared to quasi-circular inspirals.

  Fortunately, it is possible to obtain an  analytical frequency-domain
version of the above $h(t)$
%, namely $\tilde h(f)$, 
in the small eccentricity limit.
To compute such a $\tilde h(f)$, one requires 
the method 
of SPA to implement the 
required Fourier Transform.
This was essentially demonstrated at the leading order
in initial eccentricity in Ref.~\cite{KKS95} and extended to
${\cal O}(e_0^8)$ in Ref.~\cite{YABW09}.
The Fourier Transform of $h(t)$, given by Eq.~(4.29) in Ref.~\cite{YABW09},
may be written as
\begin{widetext}
\begin{align}
\tilde{h}(f) =  \mathcal{\tilde{A}}    {\left(\frac{G m \pi f}{c^3}\right)}^{-7/6}     \sum\limits_{j=1}^{10} \xi_{j}
{\left(\frac{j}{2}\right)}^{2/3}  e^{-i(\pi/4 + \Psi_j)}         ,             \label{5} 
\end{align}
\end{widetext}
where the amplitude coefficients $\mathcal{\tilde{A}}$ and $ \xi_j $ are given by 
\begin{subequations}                         \label{6}  
\begin{align}
 \mathcal{\tilde{A}} &= - {\left(\frac{5 \eta \pi}{384}\right)}^{1/2}  \frac{G^2 m^2}{c^5 D_L}  ,   \label{6.b}\\              
 \xi_j  &=  \frac{\left(1-e_t^2\right)^{7/4}}{{\left(1+\frac{73}{24}e_t^2+\frac{37}{96}e_t^4\right)}^{1/2}} \alpha_{j} e^{-i \phi_j(f/j) }    .                                      \label{6.c}                                
\end{align}
\end{subequations}
To operationalize the above expression, a number of steps are 
required.
First, the coefficients $\xi_j$ should  be Taylor expanded 
around $e_t=0$, leading to  certain
explicit expressions for $\xi_j$ in terms of $e_t, F_{\times}, F_+, \iota
$ and $\beta$. The Eqs.~(C1) of Ref.~\cite{YABW09} list such expressions for $\xi_j$ that are  
accurate to ${\cal O}(e^8)$ while choosing $ \iota= \beta=0$.
In the second step, one specifies with the help of the SPA how 
$e_t$ and $\Psi_j$  depend on the Fourier frequency $f$.
Following  Ref.~\cite{YABW09}, the expression for $\Psi_j(f)$ 
is given by
\begin{align}
\Psi_j [F(t_0)] = 2 \pi  \int_{}^{F(t_0)}     \tau' \left( j - \frac{f}{F'}   \right)    \,d{F'}  .                \label{7} 
\end{align}
where $\tau$ stands for $F / \dot {F}$.
Additionally, 
the integrals on the right hand side should be evaluated at the 
stationary point $t_0$ which is defined by $F(t_0) = f/j$.
%and the quantify $\tau \equiv F/ \dot F = \omega / \dot {\omega}$.

  A close inspection reveals that the  
 $\tau $  integrals 
can only be tackled in an approximate manner due to the GW induced evolution of $e_t$.
Therefore, the third step involves obtaining 
an approximate expression for 
$e_t$ in terms of $\omega, \omega_0$ and $e_0$. 
Subsequently, one  evaluates  the above $\tau$  integrals analytically
with the help of such an $e_t$ relation.
An approximate frequency evolution for $e_t$ is obtained by first computing the ratio
$ d \omega/d e_t = \omega\, \kappa_N (e_t) $ using Eqs.(\ref{4.b}) and (\ref{4.c}) 
for $ \dot \omega $ and $\dot{e}_t$. 
It turns out that $\kappa_N $ depends only on $e_t$ 
at the dominant quadrupolar  order.
This allows one  to write $ d \omega / \omega = \kappa_N (e_t)\, d e_t$ 
which can be integrated analytically.
The resulting expression may be written as 
$ \omega  / \omega_0 = \kappa' (e_t, e_0)$ 
and the explicit functional form of $\kappa' (e_t, e_0)$, extracted from Eq.~(62) of
Ref.~\cite{DGI},
reads
\begin{widetext}
\begin{align}
\kappa' (e_t, e_0)= 
\frac { e_0^{18/19}\,
(304 + 121\,e_0^2)^{1305/2299} }{(1-e_0^2)^{3/2}} 
\, \frac{ ( 1-e_t^2)^{3/2}}{ e_t^{18/19}\, (304 + 121\,e_t^2)^{1305/2299}}\    .   \label{7.b}
\end{align}
\end{widetext}

It is possible to invert the above expression in the limit 
 $e_t \ll 1$ to obtain 
 $e_t$ in terms of $e_0, \omega$ and $\omega_0$.
At the leading order in $e_0$, one obtains
\begin{align}
e_t &\sim e_0 \chi ^{-19/18}  + \mathcal{O}(e_0^3)  ,         \label{7.c}
\end{align}
where $\chi$ is defined as $ \omega/ \omega_0 = F/F_0$.
The  above equation  motivated Ref.~\cite{KKS95} to introduce 
an asymptotic  eccentricity invariant.
It is fairly straightforward to compute $\tau $ in terms of 
 $\omega, \omega_0$ and $e_0$ as
\begin{widetext}
\begin{align}
\tau  \sim    \frac{5~}{96~\eta~x^4}\left(\frac{G~m}{c^3}\right) \left[ 1-\frac{157 e_0^2}{24 }\chi ^{-19/9} + \mathcal{O}(e_0^4)\right]  .            \label{7.d}
\end{align}
\end{widetext}
%%%%\tau &\sim \frac{5 M}{384\ 2^{2/3} (F M)^{8/3} \pi ^{8/3}} \left[1-\frac{157 e_0^2}{24 \chi ^{19/9}} + {\cal O}(e_0^4)
The above  expression for $\tau$ allows one 
to evaluate analytically the following indefinite integral 
\begin{align}  
 2 \pi  \int     \tau'\left( j - \frac{f}{F'}   \right)    \,d{F'}  .          \label{7.e}   
\end{align}
Clearly, this integral 
will have to be evaluated at the stationary 
point $t_0$, namely 
 $j \, \dot{l }(t_0) = 2 \pi f$,
to   obtain $\Psi_j (\omega (t_0 ) )$ as noted in  Ref.~\cite{YABW09}.
This leads to the following 
 expression for $\Psi_j$,  accurate up to $ {\cal O}(e_0^2)$, 
\begin{widetext}
\begin{align}
 \Psi_j &\sim j \phi_c - 2\pi f t_c 
   - \frac{3}{128 \eta     }{  \left( \frac{G m \pi f }{c^3}    \right)   }^{-5/3} \left(\frac{j}{2}\right)^{8/3}
   \left[1-\frac{2355 e_0^2}{1462 }\chi ^{-19/9} + \mathcal{O}(e_0^4)     \right ]\,,              \label{8}
\end{align}
\end{widetext}
where $\phi_c$ and $t_c$ are the orbital phase at coalescence and the time of coalescence, respectively.
The following points are worth mentioning:
The $\chi$, appearing in the above equations for $e_t$ and $\Psi_j$,
now stands for $f/f_0$
due to the use of the stationary phase condition. 
%the $\chi$ in the above equation stands for $f/f_0$.
To ensure
that $e_t(f_0) = e_0$, one is required to rescale 
$F_0$ 
such that $F_0 \rightarrow f_0/j$.
We have verified that the above expression is indeed consistent with Eq.~(4.28) of Ref.~\cite{YABW09} that employs the chirp mass to characterize the binary.
 
   This sub-section may be summarized as follows.
The stationary phase approximation can be applied to 
compute analytically
the Fourier transform of 
the time-domain detector strain 
$h(t)$ for quadrupolar order GWs from compact binaries in
inspiraling eccentric orbits.
The resulting frequency-domain response function
 is symbolically given by Eq.~(\ref{5}).
To operationalize $\tilde h(f)$, one needs to specify the explicit functional 
dependence of $\xi_{j}$, $e_t$ and $\Psi_j$ on $f$.
The expressions for $e_t$ and $\Psi_j$ that are accurate to 
leading order in $e_0$ are given by Eqs.~(\ref{7.c}) and (\ref{8}) where 
$\chi = f/f_0$ due to the use  of the stationary-phase condition.
Additionally, we need to re-expand $\xi_{j}$ in the limit 
$e_t \ll 1$ and employ an appropriate  $e_t (f)$ 
expression to obtain the fully analytic  $\tilde h(f)$.
It is fairly straightforward  to compute higher order corrections 
in terms of $e_0$ to $e_t$ up to
${\cal O}( {e_0^7})$
  and  to extend  $\Psi_j$ to ${\cal O}( {e_0^8})$ 
as done in Ref.~\cite{YABW09}.
% and $\xi_{j}$ as
%done up to  ${\cal O}( {e_0^8})$ in Ref.~\cite{YABW09}.
In the next section, we improve  their results by
incorporating into $\Psi_j$ effects of PN-accurate eccentricity.

\subsection{Restricted $\tilde h(f)$ with 2PN order Fourier phase }   \label{sectionB}

 We begin by displaying the time-domain response function 
  for eccentric binary inspirals that incorporates 
  the first eight harmonics with quadrupolar 
order amplitudes.
The aim of this subsection, as noted earlier, is to obtain an 
analytic frequency-domain version of such a detector strain. 
Invoking Ref.~\cite{YABW09}, we write
\begin{widetext}
\begin{align} 
h(t)=    -  \frac{G m \eta }{c^2 ~D_L}x                  \sum\limits_{j=1}^{8}  \alpha_j \left(\cos{\phi_j} \cos{j l}- \sin{\phi_j \sin{j l}}  \right)\,.             \label{9}
\end{align} 
\end{widetext}
A different restriction on the harmonic index $j$ arises as our
PN-accurate Fourier phase 
will be accurate only up to ${\cal O}(e_0^6)$ at each PN order.
Similar restrictions apply while 
explicitly implementing the quantities 
$\alpha_j$ and $ \phi_j$,
given by Eqs.~(4.22) of Ref.~\cite{YABW09} in the above equation for $h(t)$.
A non-rigorous argument for restricting the harmonic 
index $j$ to six is presented towards the end of 
Sec.~\ref{sec:eT4}.
The temporally evolving $h(t)$ of Ref.~\cite{YABW09}, in principle, 
is obtained by allowing $e_t$ and $\omega$ to vary in time due to the 
quadrupolar (Newtonian) order gravitational wave emission.
However, the time evolution of the above $h(t)$ is specified 
with the help of 2PN-accurate 
differential equations for 
$\omega$ and $e_t$,
%,  extracted from Ref~\cite{ABIS}, 
%and
given by our Eqs.~(\ref{Eq:evoleqa}) and (\ref{Eq:evoleqb}) respectively.
These 2PN-accurate expressions include certain 
`instantaneous' contributions to $ d \omega/dt$ and $d e_t/dt$, given
by  Eqs.~(6.14),(6.15a),(6.15b) and (C6), Eqs.~(6.18),(6.19a), (6.19b)
and (C10) of Ref.~\cite{ABIS}, respectively. The 1.5PN order hereditary contributions to 
$ d \omega/dt$ and $d e_t/dt$ are computed with the help of 
leading order contributions to energy and angular momentum 
fluxes, given in Eqs.~(6.8) and (5.29) of Ref.~\cite{ABIS}. 
The use of above mentioned equations of  Ref.~\cite{ABIS}
in this paper implies
that we employ the harmonic gauge to obtain $h(t)$
for eccentric inspirals.
The resulting $h(t)$ models detector strain for GWs from compact binaries 
inspiraling under the influence of 2PN-accurate GW emission 
along Newtonian eccentric orbits.
In this section, we
compute
the Fourier transform of the resulting $h(t)$ analytically
while keeping terms up to ${\cal O}(e_t^6)$ at each PN order.
In contrast, the next section provides 
GW polarization states for compact binaries 
inspiraling under the influence of 2PN-accurate GW emission 
along 2PN-accurate eccentric orbits.

 We begin by listing  our main results and then  explain in detail 
how we derived them.
The  expression for $\tilde h(f)$ with 2PN level Fourier  
phase and Newtonian order amplitude reads
\begin{align}
\tilde{h}(f) =  \mathcal{\tilde{A}}    {\left(\frac{ G m \pi f}{c^3}\right)}^{-7/6}     \sum\limits_{j=1}^{8} \xi_{j}
{\left(\frac{j}{2}\right)}^{2/3}  e^{-i(\pi/4 + \Psi_j)}           ,               \label{10}
\end{align}
where the quantities $\xi_j$ are polynomials in $e_t$ 
whose coefficients are complex functions of 
$F_+, F_{\times}, \iota, \beta$
and arise from Eq.~(\ref{6.c}).
 For the present investigation,
the $\xi_j$ coefficients need 
only be accurate to ${\cal O}(e_t^6)$ due to the 
above $j$ restriction.
The main result of this section, namely, the  explicit 2PN order expression for $\Psi_j$ 
that incorporates ${\cal O}(e_0^2)$  corrections at each PN order is given by

\begin{widetext}
\begin{align}
 &\Psi_j \sim 
     j \phi_c- 2\pi f t_c   -\frac{3}{128 \eta} \left( \frac{G m \pi f}{c^3}\right)^{-5/3}  \left(\frac{j}{2}\right)^{8/3}   \left\lbrace1-\frac{2355 e_0^2}{1462 }\chi ^{-19/9}+        x \left[\frac{3715}{756}                 +\frac{55  
   }{9}\eta+      \left(      \left(-\frac{2045665}{348096}  
      \right.\right.\right. \right.\nonumber\\
   &\qquad  \left.\left.\left.          -\frac{128365
    }{12432}\eta\right)\chi ^{-19/9}                  +     \left(-\frac{2223905}{491232}+\frac{154645  }{17544}\eta\right) \chi ^{-25/9}           \right) e_0^2\right]    +  x^{3/2} \left[-16 \pi +\left(    \frac{65561 \pi }{4080 }\chi ^{-19/9}    
            \right.\right.     \nonumber\\ 
  &\qquad \left.\left.        -\frac{295945 \pi }{35088}\chi^{-28/9}     \right) e_0^2\right]    +x^2 \left[\frac{15293365}{508032}  +\frac{27145  }{504}\eta  +\frac{3085 }{72}\eta
   ^2
  +\left[     \left(-\frac{111064865}{14141952}-\frac{165068815  }{4124736}\eta                    
                       \right.  \right.\right.\nonumber\\
   &\qquad \left.\left.\left.\left. {}    -\frac{10688155 }{294624}\eta^2\right)\chi ^{-19/9}        +      \left(-\frac{5795368945}{350880768}+\frac{4917245  }{1566432}\eta+\frac{25287905 }{447552}\eta^2\right)\chi
   ^{-25/9}      +       \left(   \frac{936702035}{1485485568
   } +\frac{3062285  }{260064}\eta         \right. \right.  \right.\right.\nonumber\\
   &\qquad \left.\left.\left.\left. {}    -\frac{14251675}{631584} \eta^2    \right)\chi
   ^{-31/9}    \right] e_0^2\right]\right\rbrace\, 
    \label{11}    ,     
\end{align}
\end{widetext}
where the use of  the stationary phase condition implies that 
$\chi= f/f_0$ and $x \equiv \left( G\,m\, \omega (t_0) /c^3\right)^{2/3} $.
This ensures that  
$ x =\left [  ( G\, m/c^3) \times ( 2\,\pi f/j )\right ]^{2/3}$.
We have verified that the above expression is consistent 
with $e_0^2$ terms of Eq.~(3) in Ref.~\cite{Favata2014}.
Additionally, the following 2PN-accurate 
analytic expression for $e_t$ is required 
to specify the frequency dependence of the 
harmonic coefficients, namely
$\xi_j$:
\begin{widetext}
 \begin{align}
&e_t \sim e_0 \left\lbrace   \chi^{-19/18}  +  x\left(   \frac{2833}{2016}-\frac{197 }{72}\eta    \right)\left[ -\chi^{-19/18} +\chi^{-31/18} \right]               +  x^{3/2}\left(   \frac{377 \pi}{144}    \right)\left[- \chi^{-19/18} +\chi^{-37/18} \right]     \right.\nonumber\\
   &\qquad \left. {}                     + x^2 \left[\left(\frac{77006005}{24385536}-\frac{1143767 
   }{145152}\eta+\frac{43807 }{10368}\eta^2\right) \chi ^{-19/18}+      \left(-\frac{8025889}{4064256}+\frac{558101  }{72576}  \eta -\frac{38809
   }{5184}\eta^2\right)    \chi ^{-31/18}  \right.\right.\nonumber\\      
    &\qquad \left.\left. {}  +    \left(  -\frac{28850671}{24385536}+\frac{27565  }{145152}\eta +\frac{33811 }{10368}\eta^2\right)\chi ^{-43/18}
   \right]      \right\rbrace \,.
           \label{11.b}                        
\end{align}
\end{widetext}
Clearly, 
the above two expressions incorporate only the leading order
initial eccentricity contributions at each PN order 
and contain uncontrolled 
terms of ${\cal O}(e_0^4)$ and  ${\cal O}(e_0^3)$, respectively. 
For this paper, we have extended the above results  to obtain
2PN order expressions for 
$\Psi_j$ and $e_t$ while 
incorporating  initial eccentricity contributions 
up to $\mathcal{O}(e_0^6)$ and $\mathcal{O}(e_0^5)$, respectively,
at each PN order.
These lengthy expressions are listed as Eqs.~(\ref{appendixpsi}) and (\ref{appendixet}) in the 
Appendix~\ref{appendixA}.
This extension 
of Ref.~\cite{YABW09}  provides a
certain {\it restricted }
PN-accurate Fourier-domain 
response function for GWs from
compact binaries in  inspiraling eccentric orbits.
Our waveforms are 
 {\it restricted} as the amplitude contributions to 
 $\tilde{h}(f)$ are at Newtonian  order 
 while the Fourier phase contributions are 
2PN-accurate. 
This is the eccentric equivalent to the restricted PN waveform families
that incorporate amplitude contributions at the 
quadrupolar order and employ PN-accurate 
orbital phase evolution while modeling GWs from quasi-circular 
inspirals \cite{DIS98}.
Such waveform families are influenced by the fact that 
 the technique of {\it matched filtering}
demands  PN-accurate modeling of GW phase evolution 
while constructing inspiral search templates.

 In the following, we explain 
with intermediate steps  our approach
  to compute the 1PN extension 
 of the Newtonian order Fourier phase, available in Refs.~\cite{KKS95,YABW09}.
 This demands the extension of the   
Newtonian relation, namely $e_t = e_0\,~ \chi^{-19/18} $, to incorporate PN and higher order $e_0$ contributions 
at every PN order. We observe that 
these computations are hierarchical at each PN order.
This is because 
the 1PN-accurate 
$e_t( \chi, e_0)$ relation that incorporates ${\cal O}(e_0)$ contributions 
will be explicitly required while extending it to 
include ${\cal O}(e_0^3)$ terms at the same PN order.
In what follows, we detail our approach 
to compute the 1PN extension of the Newtonian $e_t = e_0\, \chi^{-19/18} $ relation.
 Our prescription demands the  
 computation of  1PN order expression for $d \omega/ d e_t$ with the help of 
Eqs.~(\ref{Eq:evoleqa}) and (\ref{Eq:evoleqb}) for $\dot \omega $ and $ \dot{e}_t$. 
This leads to an equation for 
$ d \omega/ \omega = \kappa_1( e_t, \omega) \, d e_t$ where
\begin{align}
\kappa_1 =  -\frac{18}{19 e_t} -\frac{3}{10108 e_t} \left(-2833+5516 \eta\right) \left(\frac{G m \omega}{c^3}\right)^{2/3}  \,. \label{12}
\end{align}
It is important to note that 
 $\omega$ terms appear only at the 1PN order.
Therefore, we employ the Newtonian accurate 
$\omega = \omega_0\, \left ( e_0/e_t \right )^{18/19}$, available in Ref.~\cite{DGI}, to 
replace $\omega$ in $\kappa_1$.      
This results in 
%leads to the following 1PN-accurate equation for $ d \omega/ \omega$:
\begin{widetext}
\begin{align}
 d \omega/ \omega \sim \left\lbrace-\frac{18}{19 e_t} -  \frac{3}{10108}   \left(\frac{e_0^{12/19}}{e_t^{31/19}}\right)
    \left(-2833+5516\eta\right)\, x_0 
    %\left(\frac{G m \omega_0}{c^3}\right)^{2/3}
  \right\rbrace de_t\,,      \label{13}
\end{align}
\end{widetext}
where  $ x_0= \left ( G\, m\, \omega_0/c^3 \right )^{2/3}$.
It is straightforward to integrate the above equation to obtain
$ \ln \omega - \ln \omega_0$ in terms of $e_t,e_0$ and $\omega_0$.
We take the exponential of the resulting expression and perform a
bivariate expansion in terms of $x_0$ and $e_t$, leading to
\begin{widetext}
\begin{align}
\omega & \sim \left\lbrace  \left(\frac{e_0}{e_t}\right)^{18/19} + x_0\left(\frac{2833-5516 }{2128}\eta   \right)     \left[  \left( \frac{e_0}{e_t}\right)^{18/19}     -       \left( \frac{e_0}{e_t}\right)^{30/19}      \right]             \right\rbrace   \omega_0  \,.
  \label{14}
\end{align}
\end{widetext}
To extract the 1PN-accurate $e_t$ expression from the above equation,
we replace the $e_t$ terms that appear at the $x_0$ level by the leading order 
 $e_t = e_0\, \chi^{-19/18} $.
It is possible to invert the resulting expression 
and obtain $e_t$ 
as a bivariate expansion in terms of $e_0$ and $x_0$.
As expected, the expansion requires that $e_0 \ll 1$ and $x_0 \ll 1$, and we 
obtain the following  $e_t$ expression:
\begin{widetext}
\begin{align}
&e_t \sim e_0 
\left\lbrace  \chi ^{-19/18}+     x_0      \left(\frac{2833}{2016}-\frac{197 }{72}\eta\right)    \left(- \chi ^{-7/18}   +  \chi
   ^{-19/18}   \right)    
\right\rbrace \,.                   \label{15}
\end{align}
We compute  $e_t$ as a bivariate expansion in terms of
the PN parameter $x$ and $e_0$ by noting that $ x/x_0 = \chi^{2/3}$.
This leads to 
% that incorporates the leading 
%order $e_0$ corrections both at the Newtonian and 1PN order
%  reads
\begin{align}
 &e_t \sim e_0 \left\lbrace   \chi^{-19/18}   +  x\left(   \frac{2833}{2016}-\frac{197 }{72}\eta    \right)\left( -\chi^{-19/18} +\chi^{-31/18} \right)              \right\rbrace         .   \label{16}
\end{align}
\end{widetext}
The hierarchical nature of these computations implies that the above expression is explicitly required during the 1PN $e_t$ computation for incorporating 
${\cal O}(e_0^3)$ terms appearing at the Newtonian and 1PN 
orders. In this paper, we pursue (and repeat) the above detailed steps to 
obtain the crucial 2PN-accurate $e_t$ as a bivariate expansion in $x$ and $e_0$.
This lengthy expression, incorporating  $ {\cal O} (e_0^5)$ corrections
at each PN order, is listed as Eq.~(\ref{appendixet}).

Let us now turn our attention to the computation of  the 1PN-accurate Fourier phase.
 We adapt Ref.~\cite{YABW09} and write PN-accurate $\Psi_j$ 
as
\begin{align}
\Psi_j [F(t_0)] = 2 \pi  \int_{}^{F(t_0)}     \tau'\left( j - \frac{f}{F'}   \right)    \,d{F'}\,,                \label{17}
\end{align}
where 
the PN approximation enters via $\tau$.
With the help of Eq.~(\ref{16}) for 1PN-accurate $e_t(\omega)$ and Eq.~(\ref{Eq:evoleqa}) for $ \dot x$,
it is straightforward to compute 1PN-accurate 
$\tau \equiv \omega / \dot {\omega}$ as
\begin{widetext}
\begin{align}
&\tau\sim         \frac{5~}{96~\eta~x^4}\left(\frac{G~m}{c^3}\right)      \left\lbrace 1 - \frac{157 e_0^2}{24}\chi^{-19/9} +  x \left[  \frac{743}{336} +\frac{11 }{4}\eta +  \left(  \left[-\frac{444781 }{24192}+\frac{30929 }{864}\eta\right]\chi^{-25/9}       +     \left[-\frac{409133}{24192}        \right.\right.\right.\right.\nonumber\\       
  &\qquad \left.\left.\left.\left. {} 
-\frac{25673  }{864}\eta\right]\chi^{-19/9}      \right) e_0^2   \right]     \right\rbrace\,.    \label{18}
\end{align}
\end{widetext}
The fact that we employ  Eq.~(\ref{16}) for $e_t$ while 
computing $\tau$ implies that 
we can only retain 
${\cal O}(e_0^2)$ contributions both at the Newtonian and 1PN orders.
It is now straightforward
to integrate analytically the indefinite integral for $\Psi_j$.
The resulting integral is evaluated at the stationary 
point $t_0$ to obtain $\Psi_j(t_0)$.
This stationary point 
 $t_0$ is again defined
to be $j \times \dot{l}(t_0) = 2 \pi f$, 
 even while invoking PN-accurate  $\tau$ expression.
 This leads to the 1PN-accurate $\Psi_j$ expression that includes 
 ${\cal O}(e_0^2)$ contributions both at the Newtonian and 1PN orders, namely
 \begin{widetext}
\begin{align}
 &\Psi_j \sim     j \phi_c- 2\pi f t_c  -   \left(\frac{3 j  }{256 \eta } \right) x^{-5/2}     \left\lbrace 1- \frac{2355 e_0^2 }{1462} \chi^{-19/9} +  x \left[ \frac{3715}{756}                 +\frac{55  
   }{9}\eta +   \left(   \left[   -\frac{2045665  }{348096}- \frac{128365 }{12432}\eta\right] \chi^{-19/9}   \right.\right.\right.\nonumber\\       
  &\qquad \left.\left.\left. {}      +  \left[ -\frac{2223905}{491232 }+\frac{154645  }{17544 }\eta\right]\chi^{-25/9}        \right)e_0^2 \right]\right\rbrace     \,,            \label{19}
 \end{align}
 \end{widetext}
where the quantities $x$ and $\chi$ will have to be evaluated 
at the stationary point.
It is straightforward, though algebraically involved, to extend
the above arguments to  2PN order while also keeping 
higher order $e_0$ contributions.
In Eq.~(\ref{appendixpsi}) of appendix ~\ref{appendixA}, we list the 2PN order $\Psi_j $ that includes all 
the ${\cal O}(e_0^6)$ contributions at every PN order.
In the remainder of this subsection, we probe 
preliminary implications of our approach.

%\subsection{Probing preliminary implications of our approach}   %\label{section3.3}

\begin{figure*}[htp]
\begin{center}
\includegraphics[width=0.8\textwidth, angle=0]{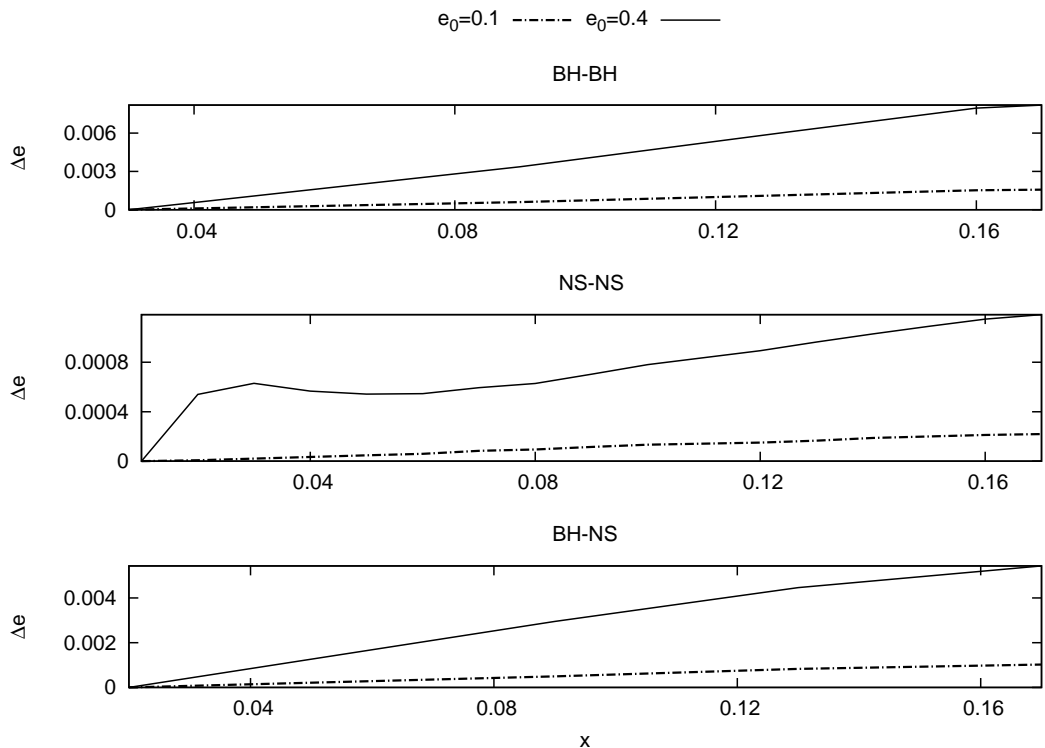}\\[0.1cm]
\end{center}
\caption{
\label{fig:1}
Plots of $\Delta e$ as a function of $x$ for three characteristic 
aLIGO compact binaries, namely BH-BH, NS-NS and BH-NS binaries (in order),
having two different initial eccentricities ($e_0=0.1, 0.4$).
The quantity $\Delta e$ stands for the difference 
between the numerical and analytical estimates for $e_t$, as discussed 
in the text.
The plots show that  $\Delta e$ is usually $\leq 2\%$ of the
initial 
$e_0$ value, even during the late inspiral ($x \sim 0.16$).
} 
%$\Delta e$ vs $x$}
%\label{fig:1}
\end{figure*}

 An obvious aspect of probing our approach should be the 
   accuracy of our bivariate expansion for the orbital eccentricity $e_t$ in terms of $x$ and $e_0$, given by Eq.~(\ref{appendixet}).
  % refer the formula in the appendix .   
  We first obtain {\it numerical} estimates $e_t^{\rm num}$ for the orbital eccentricity at certain values of 
  orbital frequency $\omega$ by numerically integrating the PN-accurate expressions 
  for $\dot{\omega}$ and $\dot{e}_t$, given by Eqs.~(\ref{Eq:evoleqa}) and (\ref{Eq:evoleqb}). 
These numerical estimates are then compared with their {\it analytic} counterparts, $e_t^{\rm ana}$, that arise from
our Eq.~(\ref{appendixet}). 
%  Let's call the values obtained by our analytic expression of eccentricity  %(Eq (A2) in the Mathematica file), $e_t^{Ana}$.
In Fig.~\ref{fig:1}, we plot the difference $\Delta e \equiv (e_t^{\rm num}-e_t^{\rm ana})$ as a function of $x$ for two different 
values of initial eccentricity, $e_0 = 0.1$ and $0.4$.
Each of the three characteristic binaries in the aLIGO frequency window is considered, i.e., 
BH-BH, NS-NS and BH-NS configurations with component masses $m_{\rm BH}=10 M_{\odot}$ and $m_{\rm NS}=1.4 M_{\odot}$, 
respectively.
Our plots reveal that the difference between the {\it exact} numerical 
%$e_t^{Num}(x)$ 
and our approximate analytical estimate for $e_t$
%$e_t^{Ana}(x)$ for  
is generally $\leq  2 \%$ of the initial value $e_0$, even during the late stage of inspiral.
% of their associated initial eccentricities even for binaries 
% having $e_0 \sim 0.4$.
This gives us confidence in employing our analytic $e_t(\omega)$
expression 
while computing the Fourier-domain response function for eccentric
inspirals.
% via Eqs.~(\ref{3.31}), (A1) and (A2).

%The $\Psi_l$ computation can be adapted to obtain 
We perform another check of our approach by computing  2PN-accurate  analytic expression for the orbital phase, 
 $\phi =\int \omega dt 
 \equiv \int \left ( \omega / {\dot \omega} ( \omega ,e_t) \right ) \, d \omega$,
   as a bivariate expansion in $x$ and $e_0$.
% This is because  of the definition of the orbital phase $\phi (x)$ as
% $ \int \omega dt 
% \equiv \int \left ( \omega / {\dot \omega} \right ) \, d \omega$.
An analytic expression for $\phi$ 
is possible as the integrand $\omega / {\dot \omega}$
becomes purely a function of $\omega$ when we employ our PN-accurate expression for $e_t(\omega)$ 
in the evolution equation for ${\dot \omega}$, as given by Eq.~(\ref{Eq:evoleqa}).
This leads to an analytic 2PN order expression for $\phi - \phi_0$, accurate to ${\cal O}(e_0^6)$, where $\phi_0$ is the initial value $\phi(x_0)$.
In what follows, we display our 2PN-accurate $\phi (x)$ that incorporates only the leading order $e_0$ contributions:
\begin{widetext}
\begin{align}
&\phi (x, e_0, F_0)   \sim  -\frac{x^{-5/2}}{32 \eta} \left\lbrace1- \frac{785}{272}e_0^{2}\chi^{-19/9}   + x\left[\frac{3715}{1008} + \frac{55}{12}\eta + \left\lbrace    \left(-\frac{2045665}{225792}-\frac{128365}{8064}\eta \right)\chi^{-19/9}         \right.\right.\right.\nonumber\\
 &\qquad \left.\left.\left.{} 
+              \left(-\frac{2223905}{274176}    +\frac{154645}{9792}\eta \right)\chi^{-25/9}   \right\rbrace e_0^{2}  \right]   + x^{3/2}\left[ -10\pi + \left(  \frac{65561 \pi}{2880 }\chi^{-19/9} -\frac{295 945 \pi}{19584 }\chi^{-28/9}    \right) e_0^2\right]
\right.\nonumber\\
&\qquad  
+ x^2 \left[ 
 \frac{15 293 365}{1016064}+ \frac{27145}{ 1008}\eta + \frac{  3085}{144}\eta^2                           + \left\lbrace              \left(- \frac{111064865}{ 10948608}-  \frac{165068815}{ 3193344}\eta - \frac{10 688 155}{228096}\eta^2                            \right)\chi^{-19/9}  \right.\right.\nonumber\\  
 &\qquad \left.\left.\left.{}                        
      +\left( -\frac{5 795 368 945}{227598336} + \frac{4 917 245 }{ 1016064 }\eta + \frac{25 287 905}{290304}\eta^2                            \right)\chi^{-25/9}     +    \left( \frac{936702035}{829108224}+ \frac{3062285}{145152}\eta - \frac{14 251 675 }{352512
}\eta^2                            \right)        \right.\right.\right.\nonumber\\  
 &\qquad \left.\left.\left.{}
\chi^{-31/9}        \right\rbrace   e_0^2 
\right]   \right\rbrace  \,.
       \label{20}
\end{align}     
\end{widetext}  
% where $\phi_0$ is the value of orbital phase at $x_0$.
 The 2PN-accurate expression for $\phi$
 that incorporates eccentricity contributions accurate up to
 ${\cal O}(e_0^6)$ is listed in as Eq.~(\ref{appendixPhi}) in the appendix.

\begin{table*}[htp]
\begin{center}
\centering
\begin{tabular}{|p{2.0cm}|p{1.8cm}|p{1.8cm}|p{1.8cm}|p{1.8cm}|p{1.8cm}|p{1.8cm}|p{1.8cm}|}
\hline 
\textbf{$e_0$} & \textbf{0} & \textbf{0.001} & \textbf{0.01} & \textbf{0.1} & \textbf{0.3} & \textbf{0.4} & \textbf{0.5}\tabularnewline
\hline 
\hline 
\multicolumn{8}{|c|}{\textbf{$m_1=10 M_{\odot}$; $m_2=10 M_{\odot}$}}\tabularnewline
\hline 
\small{eT4} & 607.72 & 607.76 & 608.89 & 590.16 & 453.58
 & 353.79 & 249.27 \tabularnewline
\hline 
\small{2PN analytic} & 613.88

(-1.01\%) & 613.88

(-1.00\%) & 613.69

(-0.78\%) & 594.75

(-0.77\%) & 456.38

(-0.61\%) & 354.79

(-0.28\%) & 247.03

(0.89\%)\tabularnewline
\hline 
\small{PC + 2PN \newline circular} & 613.88

(-1.01\%) & 613.88

(-1.00\%) & 613.71

(-0.79\%) & 596.59

(-1.08\%) & 470.65

(-3.76\%) & 377.04

(-6.57\%) & 276.33

(-10.85\%)\tabularnewline
\hline 
\small{2PN EPC} & 613.88

(-1.01\%) & 613.88

(-1.00\%) & 613.78

(-0.80\%) & 603.85

(-2.31\%) & 521.77

(-15.03\%) & 449.74

(-27.12\%) & 361.90

(-45.18\%)\tabularnewline
\hline 
\multicolumn{8}{|c|}{\textbf{$m_1=1.4 M_{\odot}$; $m_2=1.4 M_{\odot}$}}\tabularnewline
\hline 
\small{eT4} & 16262.75 & 16262.70 & 16257.98 & 15790.74 & 12358.47 & 9818.22 & 7114.64\tabularnewline
\hline 
\small{2PN analytic}& 16274.97

(-0.07\%) & 16274.92

(-0.07\%) & 16270.20

(-0.07\%) & 15802.74

(-0.07\%) & 12368.10

(-0.07\%) & 9821.08

(-0.02\%) & 7088.08

(0.37\%)\tabularnewline
\hline 
\small{PC + 2PN \newline circular} & 16274.97

(-0.07\%) & 16274.92

(-0.07\%) & 16270.34

(-0.07\%) & 15816.86

(-0.16\%) & 12480.40

(-0.98\%) & 10000.20

(-1.85\%) & 7332.09

(-3.05\%)\tabularnewline
\hline 
\small{2PN EPC} & 16274.97

(-0.07\%) & 16274.94

(-0.07\%) & 16272.36

(-0.08\%) & 16012.69

(-1.40\%) & 13866.57

(-12.20\%) & 11983.00

(-22.04\%) & 9685.34

(-36.13\%)\tabularnewline
\hline 
\multicolumn{8}{|c|}{\textbf{$m_1=10 M_{\odot}$; $m_2=1.4 M_{\odot}$}}\tabularnewline
\hline 
\small{eT4} & 3605.67 & 3606.06 & 3607.44 & 3499.05 & 2706.34 & 2124.47 & 1511.46\tabularnewline
\hline 
\small{2PN analytic}& 3618.89

(-0.36\%) & 3618.88

(-0.35\%) & 3617.78

(-0.28\%) & 3508.86

(-0.28\%) & 2711.70

(-0.19\%) & 2124.69

(-0.01\%) & 1499.74

(0.77\%)\tabularnewline
\hline 
\small{ PC + 2PN \newline circular} & 3618.90

(-0.36\%) & 3618.89

(-0.35\%) & 3617.86

(-0.28\%) & 3516.49

(-0.49\%) & 2770.61

(-2.37\%) & 2216.16

(-4.31\%) & 1619.70

(-7.16\%)\tabularnewline
\hline 
\small{2PN EPC} & 3618.90

(-0.36\%) & 3618.89

(-0.35\%) & 3618.31

(-0.30\%) & 3560.04

(-1.74\%) & 3078.48

(-13.75\%) & 2655.95

(-25.01\%) & 2140.77

(-41.63\%)\tabularnewline
\hline 
\end{tabular}
\end{center}
\caption{Four different estimates for the accumulated number of GW cycles
associated with the $l=2$ harmonic
of eccentric compact binary inspirals 
in a frequency window, defined 
 by the earlier specified $x_{\rm low}$ and $x_{\rm high}$ values.
The four 2PN order $\mathcal{N}$ estimates are based on
one purely numerical (eT4)
and three different analytic expressions for $\phi$, as detailed in the text. 
The displayed 
fractional differences in $\mathcal{N}$ probe 
how the three analytic  $\mathcal{N}$ estimates differ from their 
purely numerical counterpart.
These fractional differences, displayed in the parentheses, are 
 computed by evaluating
$ \left ( [ \mathcal{N}_{\rm num}-\mathcal{N}_{\rm ana} ] / \mathcal{N}_{\rm num} \right ) \times 100\% $.
%We let initial orbital frequency to be $10\, \pi$ Hz and let $x_{\rm max} = %x(F_{max})$ as explained in the text. 
}
\label{table:1}
\end{table*}

 With this input,
we pursue a check on our  PN-accurate 
 $\Psi_j$ by
 computing the accumulated number of GW cycles
associated with the $j=2$ harmonic in the frequency 
window, specified by 
$x_{\rm low} ={ \left(   {G\, m\, \pi\, 10 } / {c^3}  \right) }^{2/3}$
and $x_{\rm high}= 1/6$.
%in the aLIGO frequency window.
%$\phi (x)$ expression allows us to perform 
% another check on our prescription to compute PN-accurate $e_t$ and 
% the resulting $\Psi_l$.
% This check involves 
% computing the accumulated number of GW cycles
%associated with the $l=2$ harmonic in the aLIGO frequency window.
We compute and compare 
four different estimates for 
$\mathcal{N} = \left ( \phi_{\rm max} - \phi_{\rm min} \right )/\pi$, where $\phi_{\rm max}$ and $\phi_{\rm min}$ 
are the values of the orbital phase at the
initial and final values of the $x$ parameter.
These four $\mathcal{N} $ estimates are evaluated for each of the classical aLIGO binaries 
while choosing a number
of $e_0$ values.
% $x_{\rm min}$ and $x_{\rm max}$ values for typical aLIGO binaries having %a number of $e_0$ values.
%$x_{\rm max} =x(F=F_{max}) $ and 
%  $x_{\rm min} = x(F_0=5 Hz)$ respectively where $F_{max}=6^{-3/2}c^3/(G m %\pi)$ (see Ref. \cite{faye}). 
The first estimate for $\mathcal{N}$ arises from
our eccentric extension of the circular {\texttt{TaylorT4}} approximant at 2PN order, denoted by `eT4' and
detailed in the next section.
This estimate may be treated to be {\it exact} in $e_t$ since we do not perform any small eccentricity expansion in our time domain approximant.
We developed this approximant with the aim to improve certain computational aspects of 
  the $x$-model which has been validated against a particular numerical relativity 
waveform for the eccentric inspiral of an equal-mass binary \cite{Hinder09}.
 The second estimate for $\mathcal{N}$ is obtained by employing 
our 2PN order (and ${\cal O}(e_0^6)$ accurate) expression for $\phi$, namely Eq.~(\ref{appendixPhi}).
In Table \ref{table:1}, we list these two estimates for $\mathcal{N}$
in the first two rows 
while considering the usual BH-BH, NS-NS and BH-NS binaries 
with initial eccentricities $e_0= 0, 10^{-3}, 10^{-2}, 0.1, 0.3, 0.4$ and $0.5$.
For each $(m_1, m_2,e_0)$ configuration, the fractional difference between the {\it numerical} and our {\it analytic} 
estimate, namely $\left ( \mathcal{N}_{\rm num} - \mathcal{N}_{\rm ana} \right ) / \mathcal{N}_{\rm num} \times 100 \%$, 
is displayed in the parentheses.

A comparison of the evaluated numbers reveals that our 2PN-accurate analytic prescription for $\phi$ 
slightly overestimates the accumulated number of GW cycles
compared to the numerical estimate for the same $(m_1, m_2,e_0)$ configuration.
However, our analytic ${\cal N}$ estimates are 
 fairly close to their numerical counterparts even for binaries having moderate initial eccentricity $e_0 \sim 0.4$.
Note that binaries with tiny initial orbital eccentricities 
exhibit higher fractional differences between the numerical and analytic estimates.
 This can be attributed to the fact that 
various time-domain PN-accurate quasi-circular inspiral template families indeed provide
slightly different $\mathcal{N}$ estimates \cite{Boyle_T4,BIOPS}.
We observe sign reversals for the quantity in parentheses
 at $e_0 \sim 0.5$
when considering our eccentric time and frequency-domain approximants. This may be treated as a pointer to the 
range of applicability of 
our frequency-domain templates.

Additionally, we computed two other $\mathcal{N}$ estimates that are 
 based on two different approaches to obtain an {\it analytic} expression for $\phi$.
These numbers are also displayed in
Table \ref{table:1} along with their fractional differences with respect to 
their numerical counterparts.
%The {\it  Newtonian PC + 2PN circular } $\mathcal{N}$ estimate 
%employs an analytic expression for $\phi$ 
%that is based on the Newtonian post-circular approach of %Ref.~\cite{YABW09}.
%To obtain {\it Newtonian PC + 2PN circular } 
The {PC + 2PN circular} estimate for
$\mathcal{N}$ is based on the Newtonian order `post-circular' $\phi $ of Ref.~\cite{YABW09}, supplemented by the
 1PN, 1.5PN and 2PN circular contributions to $\phi$
available in Ref.~\cite{BDI}.
% to the Newtonian order Ref.~\cite{YABW09}
%based $\phi$ that is accurate to ${\cal O}(e_0^6)$.
In contrast,
the 2PN order {EPC} estimate for $\mathcal{N}$ is based on Ref.~\cite{H14}.
This estimate  employs a certain 2PN order analytic $\phi$ 
that includes eccentricity corrections accurate up to ${\cal O}(e_0^6)$.
Here, the PN order eccentricity contributions to the phase are 
computed from the standard 2PN-accurate 
circular version of $\phi$ by employing a certain modified velocity
function $v_{ecc}$, given by Eq.~(13) of Ref.~\cite{H14}.
It should be noted that the resulting $\phi (x,e_0)$ is not a consistent
PN expansion to 2PN order of the PC approach (this aspect of the EPC
approach was noted 
in Ref.~\cite{H14}).
 The numbers listed in Table \ref{table:1} reveal that the $\mathcal{N}$ estimates obtained with the EPC approach 
differ substantially from the purely numerical estimates
even for binaries with initial eccentricity $e_0 \sim 0.1$.
This is troubling since
our $\mathcal{N}_{\rm num}$ estimates, as noted above,
are based on an improved version of the numerical relativity calibrated $x$-model for eccentric 
inspirals, which treats eccentricity contributions in an exact manner.
In comparison, the numbers arising from a modified PC approach which incorporates 
only circular contributions to $\phi$ at PN orders are closer to our `eT4' based
$ \mathcal{N}$ estimates even for $e_0 \sim 0.1$.
Observe that all analytic $\phi$ based $\mathcal{N}$ estimates 
are close to each other for 
tiny residual eccentricities like $e_0=10^{-3}$ or $10^{-2}$.
However, we glean from additional evaluations 
that our 2PN-accurate analytic $\mathcal{N}$ estimates are 
comparatively closer to their `eT4' counterparts 
even for compact binaries with non-negligible initial 
eccentricities like $e_0 = 0.3$.
This observation and 
the various estimates of Table~\ref{table:1}
indicate, in our opinion, the need to incorporate 
eccentricity evolution contributions in a PN-accurate and 
consistent manner while 
computing $\phi$ and the associated $\mathcal{N}$ estimates.

Note that the integral that defines the orbital phase $\phi$ is key to obtain the Fourier phase
$\Psi_j$ in the SPA, as evident from  Eqs.~(4.7) and (4.8) in Ref.~\cite{YABW09}. This suggests that
one should also incorporate 
eccentricity evolution in a PN-accurate manner while 
computing the PN-accurate Fourier phase.
Therefore, our computation 
%our consistent PN extension of the PC approach
should be useful to construct accurate and computationally efficient Fourier-domain 
search templates for compact binaries in inspiraling 
eccentric orbits. Clearly, further extension 
and investigation will be required to substantiate this 
statement.

Let us again summarize our main result. The fully analytic frequency-domain response function, applicable for 
 GW data analysis investigations, with Newtonian order amplitude 
 and 2PN order Fourier phase $\Psi_j$
% Fourier transform (accurate up to Newtonian level in amplitude and 2PN level %in the phase) 
 is given by Eq. (\ref{10}), where
the quantities  $\mathcal{A}$ and $\xi_{j}$ are given by Eqs. (\ref{6.b}) and (\ref{6.c}).
Clearly, we need to
perform an expansion around small eccentricity $e_t$ while explicitly using the quantities $\xi_{j}$.
 The associated $e_t$ and $\Psi_j$ expressions 
  % appearing in either $\tilde{h}(f)$ or $\xi_{\ell}$ 
  are given analytically 
  by Eqs. (\ref{appendixet}) and (\ref{appendixpsi}).
In the next section, we explain the approach that allowed us 
to provide  exact numerical estimates for the accumulated number of GW cycles.

\section{Incorporating orbital eccentricity into the TaylorT4 approximant }
\label{sec:eT4}

In this section, we present an accurate and efficient prescription to incorporate orbital eccentricity 
into the quasi-circular time-domain 2PN-accurate {\texttt{TaylorT4}} approximant. 
The reasons for focusing only on the 
{\texttt{TaylorT4}} approximant are the following.
We observe that the circular {\texttt{TaylorT1}}
approximant provides a differential 
 equation for $x$ which is essentially a ratio of polynomials in $x$ \cite{BIOPS}.
Therefore, its straightforward eccentric extension 
requires us to 
expand the differential equations for 
 $x$ and $e_t$, given by Eqs.~\ref{Eq:evoleq},  as bivariate expansions 
 in terms of $x$ and $e_t$. The 
 resulting expressions can be used to obtain differential equations for 
 $x$ and $e_t$ as ratios of polynomials in $x$ and $e_t$.
 Clearly, this is inconsistent with our efforts to include 
 $e_t$ in an exact manner.
 Technically, it is also possible to express 
   the differential equations for 
 $x$ and $e_t$ as ratios of polynomials in $x$ while keeping $e_t$ contributions 
 rather exact in $e_t$. We have worked out such a model
  to 1PN order, 
 and the resulting approximant turned out to be noticeably slower (computationally)
  than its  eccentric  {\texttt{TaylorT4}} counterpart. In our opinion,
 this version is computationally slower mainly due to the presence of 
 $1/e_t$ terms in the differential equations for $e_t$.
 Such terms are
 rather unavoidable due to the Newtonian-accurate 
 $e_t^2 = 1 + 2\, E\, J^2$ expression,
 where $E$ and $J$ stand for the reduced 
 orbital energy and angular momentum, respectively.
 The above two observations prompted us not to 
 pursue  {\texttt{TaylorT1}} approximant 
 while including the effects of orbital $e_t$.
 We note that the eccentric versions of both 
 {\texttt{TaylorT2}} and {\texttt{TaylorT3}} approximants
 will also force  us to treat $e_t$ in an approximate manner. 
 This restriction  is required  
 to obtain analytic expressions for $\left [ \phi(\omega), 
 e_t( \omega), t(\omega) \right ]$ 
 and $\left [ \phi(t), \omega(t), e_t(t) \right ]$ that are crucial to obtain 
 eccentric versions of 
 the circular {\texttt{TaylorT2}} and {\texttt{TaylorT3}} approximants,
 respectively.
Therefore, straightforward eccentric versions of both 
{\texttt{TaylorT2}} and {\texttt{TaylorT3}} approximants
are also in conflict with our desire to treat $e_t$ in an exact manner.
In what follows, we  briefly sketch how we 
adapt the GW phasing formalism of Ref.~\cite{DGI} to
include $e_t$ effects in an accurate and exact manner 
into the circular {\texttt{TaylorT4}} approximant.
The salient features of our approach  and 
preliminary data analysis implications via certain match estimates are also presented.

\subsection{GW phasing for compact binaries in inspiraling 2PN-accurate eccentric orbits}

We begin by listing 
the dominant quadrupolar contributions to the two independent GW polarization 
states, $h_{+}\big|_{\rm Q}(t)$ and $h_{\times}\big|_{\rm Q}(t)$,
associated with a $(m, \eta)$ 
compact binary 
 at a luminosity distance $D_L$ from the observer:

\begin{subequations}
\label{Eq:hpx_ecc}
 \begin{align}
h_{+}(r,\phi,\dot{r},\dot{\phi}) \big|_{\rm Q}
&=-\frac{G m \eta}{c^4 D_L}
\bigg\{
(1 + C^2)
\bigg[
\bigg( \frac{G m}{r} + r^2 \dot{\phi}^2 
\nonumber
\\
& \quad
- \dot{r}^2 \bigg) \cos 2 \phi
+ 2 \dot{r} r \dot{\phi} \sin 2 \phi
\bigg]
\nonumber
\\
& \quad
+ S^2 \bigg[
\frac{G m}{r} - r^2 \dot{\phi}^2 - \dot{r}^2
\bigg]
\bigg\}
\,,
\\
h_{\times}(r,\phi,\dot{r},\dot{\phi}) \big|_{\rm Q}
&=-\frac{2 G m \eta C}{c^4 D_L}
\bigg[
\bigg( \frac{G m}{r} + r^2 \dot{\phi}^2 
\nonumber
\\
& \quad
- \dot{r}^2 \bigg) \sin 2 \phi
- 2 \dot{r} r \dot{\phi} \cos 2 \phi
\bigg]
\,,
\end{align}
\end{subequations}
where 
%$\eta = \mu / m = m_1 m_2 / (m_1 + m_2)^2$ is the symmetric mass ratio and
 $C$ and $S$ 
denote $\cos i$ and $\sin i$, respectively, with $i$ being
the inclination of the orbital plane with respect to the plane of the sky
  (see, e.g., Refs.~\cite{GI02,DGI}).
The dynamical variables  $r$, $\dot{r}$, $\phi$ and $\dot{\phi}$ define 
the polar coordinates of the relative orbital separation vector and their time derivatives.

 The GW phasing formalism, developed in Ref.~\cite{DGI}, provides an efficient way of 
implementing both the conservative and reactive contributions to the temporal evolution for 
these  dynamical variables $\{r, \dot{r}, \phi, \dot{\phi}\}$ appearing 
in Eqs.~(\ref{Eq:hpx_ecc}). 
The approach involves splitting the binary dynamics 
 into  conservative and  dissipative parts, 
with the latter first entering 
the compact binary dynamics 
at the 2.5PN (absolute) order.
The 2PN-accurate conservative part of the orbital dynamics is integrable 
and admits an analytic solution, namely a Keplerian-type
parametric solution as detailed in Ref.~\cite{MGS}.
The existence of such a 2PN-accurate Keplerian-type parametric solution 
allows us to express the  radial and angular
parts of the orbital dynamics as 
\begin{align}
\label{Eq:dynvargen}
r(t) &= r~(u(l), {\cal E}, {\cal J} ) ,&\dot r(t) = \dot r~(u(l), {\cal E}, {\cal J} )\,,  \\
\phi(t) &= \lambda + W(u(l), {\cal E}, {\cal J}),&\dot \phi (t) = \dot \phi~(u(l), {\cal E}, {\cal J}) \,,
\end{align}
where $u$ and $l$ are the eccentric and mean anomalies 
of the 
Keplerian parametrization, while ${\cal E}$ and 
${\cal J}$ stand for the orbital energy and the angular momentum, respectively. 
The split of the angular variable $\phi$  explicitly incorporates the effect of periastron 
advance. This is particularly useful while constructing the frequency 
spectrum associated with $h_{+,\times}\big|_{\rm Q}(t)$ of binaries in inspiraling 
eccentric orbits (see Ref.~\cite{TG06} for details).
The angular variable $W$ is $2\pi$-periodic in $u$ and analytically models 
orbital time scale variations in $\phi$.
The remaining two
angular-type variables 
$l$ and $\lambda$ are defined to be
\begin{align}
\label{Eq:l_lambda}
l & \equiv n ( t- t_0) + c_l\,,~~\lambda & \equiv ( 1 + k ) n ( t- t_0) + c_{\lambda}\,,
\end{align}
where the constants $t_0$, $c_l$ and $ c_{\lambda}$ refer to an initial instant and the respective 
$l$ and ${\lambda}$ values at $t= t_0$.
The parameter $n$ is usually referred to as the mean motion while $k$ 
measures the  periastron advance
in the time interval $P = 2\,\pi/n$. It should be noted that the PN-accurate expressions 
for $n$ and $k$ in terms of 
${\cal E}$ and ${\cal J}$ are gauge-invariant quantities \cite{DS88,MGS}.

   An additional equation is required to specify how $u$ varies with $l$ 
and therefore to model explicitly the  temporal evolution of our  
dynamical variables $\{r, \dot{r}, \phi, \dot{\phi}\}$ .
This is done by solving the following 
 2PN-accurate Kepler equation (KE) to find $u(l)$.
 At 2PN order,  the KE 
can be symbolically expressed as
\begin{align}
\label{Eq_2:KE_0123}
l &=  u - e_t\, \sin u + l_{\rm 2} (u, {\cal E}, {\cal J})\,,
\end{align}
where $l_{\rm 2}$ denotes the 2PN corrections to the usual Newtonian KE,
namely $l = u -e_t\, \sin u$. In above equation, $e_t$ stands for a
 certain `time-eccentricity' parameter of the 
Keplerian-type parametric solution to the PN-accurate orbital dynamics.
A few comments are in order before we explain the details of the GW phasing formalism. 
It is customary to employ 
$\omega = n\, ( 1 +k)$ and $e_t$ to characterize PN-accurate 
eccentric orbits instead of  ${\cal E} $ and $ {\cal J}$.
This  ensures that $\phi$ becomes the required  $\omega ( t- t_0)$ in the circular limit~\cite{ABIS}.
Additionally, Ref.~\cite{Hinder09} showed that inspiral waveforms that employ 
$\omega$ and $e_t$ are in better agreement with their numerical relativity counterparts
while considering equal mass eccentric inspirals.
These considerations influenced us to employ $x$ and $e_t$ to characterize PN-accurate 
eccentric orbits as done in Ref.~\cite{Hinder09,ABIS,ABIQ}.
This implies that  our compact binary, evolving under the influence of 2PN-accurate 
binary dynamics, is fully specified by four initial parameters, namely the values 
of $x,e_t, c_l$ and $c_{\lambda}$ at the initial epoch.
 
 The effects of the dominant (quadrupolar) order GW 
emission enters the binary dynamics at 2.5PN order.
 An improved `method of variation of constants'  was developed in Ref.~\cite{DGI}
 to include the effects of GW emission on the conservative 
 2PN-accurate dynamics of compact binaries 
in precessing eccentric orbits.
This is implemented by demanding that the fully 2.5PN-accurate binary dynamics
 preserves the same functional form for the dynamical variables  $\{r, \dot{r}, \phi, \dot{\phi}\}$.
 However, the constants of the conservative dynamics are allowed to vary in time.
The equations governing the temporal evolutions of these `constants' are
given by Eqs.~(35) in Ref.~\cite{DGI} and depend, as expected,
on the reactive contributions to the binary dynamics.
Therefore, this approach allowed Ref.~\cite{DGI} to describe the orbital evolution of 
eccentric binaries under the influence of fully 2.5PN-accurate orbital motion
in a semi-analytic manner.

 It was demonstrated in Ref.~\cite{DGI} that  the temporal variations of the four constants of
integration can be decomposed as a combination of a slow drift and fast oscillations.
We may write such variations  
symbolically as
\begin{align}
\label{phasing_eq:18}
c_\alpha (l) & =
\bar{c}_\alpha (l) + \tilde{c}_\alpha (l)
\,,
\end{align}
where the subscript $\alpha$ stands for  one of the four `constants of motion'. ( In Ref.~\cite{DGI}, these constants were
chosen to be 
 ${\cal E}, {\cal J}, c_l $ and $c_{\lambda}$.)
In the above equation, $\bar{c}_\alpha (l)$ denotes the slow (secular) drift,
which accumulates over the radiation reaction  time scale to induce large changes in $c_\alpha (l)$.
The fast (periodic) oscillations denoted by $\tilde{c}_\alpha (l)$ are orbital time scale 
variations in $c_\alpha (l)$.
It turned out that 
the effects of  such rapidly oscillating 
contributions are of substantially smaller magnitudes than those associated with the slow drift
even while including higher order radiation reaction effects \cite{DGI,KG06}.
Therefore, in the present work we consider only the secular time scale variations $ \bar{c}_\alpha (t)$ 
in the orbital dynamics.

  Detailed computations reveal that 
$ d \bar {c_l} /dt = d \bar {c_{\lambda}} /dt  \equiv 0$ even while
including higher order radiation reaction terms in the orbital dynamics \cite{DGI,KG06}.
Additionally, the differential equations for the other two secular variables are 
identical to the PN-accurate  expressions for 
far-zone energy and angular momentum fluxes, provided 
$ {\cal E} $ and $  {\cal J}$ were used to characterize the orbit.
This is a highly desirable result as 
the far-zone energy and angular momentum fluxes are available to 
higher PN orders, compared to the 1PN-accurate expressions for the reactive contributions 
to the orbital dynamics.
This allowed Ref.~\cite{DGI} to model orbital dynamics of compact binaries 
inspiraling under the influence of GW emission at the 2PN order 
while moving along 2PN-accurate eccentric orbits.
It should be noted that this approach does not require one to use 
$ {\cal E} $ and $  {\cal J}$ to specify the orbit.
It is indeed possible to use, for example, $\omega$ (or $x$) and $e_t$ to specify our PN-accurate 
eccentric orbit along with $c_l$ and $c_{\lambda}$.
The relevant differential equations for $\bar \omega$ and $\bar e_t$ 
are computed with the help of the `balance' arguments.
This involves  invoking the  2PN-accurate expressions for $\omega$ and $e_t$ 
in terms of  $ {\cal E} $ and $  {\cal J}$ and employing  the balance arguments that equate 
the time derivatives of   $ {\cal E} $ and $ {\cal J}$ to the 2PN-accurate 
far-zone energy and angular momentum fluxes.
As a result, one finds 2PN-accurate expressions for 
$ d \bar x/dt$ and $ d \bar e_t/dt$ that incorporate the 
secular effects of GW emission at 2PN order.
In what follows, we explain how we adapt and improve (numerically) the GW phasing approach 
of  Ref.~\cite{DGI} by employing  $x,e_t, c_l$ and $c_{\lambda}$ to characterize 
2PN-accurate eccentric orbits.
Therefore, our approach parallels the $x$-model in some aspects, and we will highlight the differences
between the two models subsequently.

 Our approach  to obtain $h_{+,\times} \big|_{\rm Q}(t)$ for compact binaries 
 that are under the influence of fully 2PN-accurate description in both 
 conservative and reactive dynamics requires certain PN-accurate parametric expressions.
 These expressions provide the 2PN-accurate 
 conservative dynamics for the variables that appear in the 
$h_{+,\times} \big|_{\rm Q}(t)$. 
We list  in a partially symbolic manner the 2PN-accurate 
equations for $\{r, \dot{r}, \phi, \dot{\phi}\}$ in terms of $x$, $e_t$ and $u$ as 
\begin{subequations}
\label{Eq:dynvar}
\begin{align}
\frac{ \dot{r} }{c} 
 &= \frac{\sqrt{x} \, e_t \, \sin{u}  }{1 - e_t\,\cos{u}} \,
		\biggl \{   	1
			+ 	\dot{r}^{\rm 1PN}(\eta,\,e_t)\,x \nonumber \\
			&+	\dot{r}^{\rm 2PN}(\eta,\,e_t,\,u)\,x^2
  \biggr \}\,, \\
%-------------------------------------------------------------------------------------------------------------------
\frac{ r\, \dot{\phi}}{c}
 &=	\frac{\sqrt{1 - e_t^2}\,\sqrt{x}}{1 - e_t\,\cos{u}}
	 \biggl \{  	1
		+ 	r^{\rm 1PN}(\eta,\,e_t,\,u)\,x \nonumber \\
		&+	r^{\rm 2PN}(\eta,\,e_t,\,u)\,x^2
\biggr\} \times
	 \biggl \{  	1 
		+ 	\dot{\phi}^{\rm 1PN}(\eta,\,e_t,\,u)\,x \nonumber \\
		&+	\dot{\phi}^{\rm 2PN}(\eta,\,e_t,\,u)\,x^2
\biggr\} \,, \\
%-------------------------------------------------------------------------------------------------------------------
\frac{G\, m}{c^2\,r}
 &= \frac{x}{1 - e_t\,\cos{u}} \,
	 \biggl\{	1
		+	r^{\rm 1PN}(\eta,\,e_t,\,u)\,x \nonumber \\
		&+	r^{\rm 2PN}(\eta,\,e_t,\,u)\,x^2
\biggr\}^{-1}\,, \\
%-------------------------------------------------------------------------------------------------------------------
\phi
&= \lambda + W(\eta,\, x,\, e_t,\, u) \,,
\end{align}
\end{subequations}
where the  explicit functional forms for the various PN contributions like $\dot{r}^{\rm 1PN/2PN}$ are provided 
in appendix \ref{appendixB}.
These expressions can be easily obtained from Ref.~\cite{KG06} while using the 2PN-accurate 
relation between $\omega$ and $n$ as given in Ref.~\cite{ABIS}. 

  The periodic contributions to the angular variable, given by $W(x,e_t,u)$, require 
  additional considerations.
Following Ref.~\cite{DGI}, we write 
\begin{align}
\label{Eq:W_2PN}
W &=
	(v - u) + e_t \, \sin u \nonumber \\
	&+ W^{\rm 1PN}(\eta,\,e_t,\,u)\,x + W^{\rm 2PN}(\eta,\,e_t,\,u)\,x^2\,,
\end{align}
where the explicit expressions for $W^{\rm 1PN/2PN}$ are also listed 
in appendix \ref{appendixB}.
We employ the following exact relation for the $(v-u)$ part of $W$ \cite{KG06}: 

\begin{align}
\label{Eq:vminu_2PN}
v - u &=
	2\, \tan^{-1}
	\left[
	\frac{ \beta_{\phi}\, \sin{u} }{ 1 - \beta_{\phi}\, \cos{u} }
	\right]\,,
\end{align}
where  $\beta_{\phi} = ( 1 - \sqrt{ 1 - e_\phi^2 } ) / e_\phi$
and $e_{\phi}$ stands for a certain `angular eccentricity' parameter of the PN-accurate 
Keplerian-type parametric solution.
The higher-order PN corrections enter while connecting $e_{\phi}$ to $e_t$ 
with the help of relevant expressions available in Ref.~\cite{MGS}.
It is  fairly straightforward to express $\beta_{\phi}$ in terms of $x$ and $e_t$ at 2PN 
order as 
\begin{align}
\label{Eq:beta_2PN}
\beta_{\phi} &=
	      \frac{1 - \sqrt{1 - e_t^2}}{e_t} \nonumber \\
	      &+ \beta_{\phi}^{\rm 1PN}(\eta,\,e_t)\,x
	      + \beta_{\phi}^{\rm 2PN}(\eta,\,e_t)\,x^2\,,
\end{align}
where the explicit expressions for the PN contributions are again listed in appendix \ref{appendixB}.
In practice, we use the 2PN-accurate expression for $\beta_{\phi}$ while 
evaluating $v-u$ terms appearing in various PN orders of $W$.

  To obtain the temporal evolution of these dynamical variables which are explicit functions 
  of $u$, we solve the following 2PN-accurate Kepler equation to connect $u$ and $l$.
  In terms of $x$ and $e_t$, the 2PN-accurate KE reads 
\begin{align}
 \label{Eq:KE_2PN}
 l &= u - e_t\,\sin{u} + x^2\,\biggl \{\frac{(15 - 6\,\eta)\,(v - u)}{2\,\sqrt{1-e_t^2}} \nonumber \\
      &+ \frac{\eta\,(15 - \eta)\,e_t\,\sin{u}}{8\,(1 - e_t\,\sin{u})}\biggr \}\,,
\end{align}
where the $v-u$ term appearing at the 2PN order on the right hand side of Eq.~(\ref{Eq:KE_2PN}) 
is evaluated using the above prescription involving $\beta_{\phi}$.

To solve the 2PN-accurate KE, we employ a modified version of Mikkola's method for solving
the classical KE, as introduced in Ref.~\cite{SM87}. This computationally inexpensive 
root-finding method involves the solution of a cubic polynomial and a subsequent 
fourth-order iteration to improve on the initial guess. Mikkola's solution is valid 
for all $l$ and for $0 \leq e_t \leq 1$ (see Ref.~\cite{TG06} for details). To solve KE at the 2PN-order, 
we first apply Mikkola's method to the relation  $l = u - e_t\,\sin(u)$ 
to obtain a certain `Newtonian' accurate $u$ value for a given $l$.
This solution is employed to determine the 2PN corrections to the usual KE, namely $l_2\,(u, x, e_t)$.
In other words, the first use of Mikkola's method allows us to obtain the temporal variation
of $l_2\,(u, x, e_t)$ appearing on the right-hand side of 
Eq.~(\ref{Eq:KE_2PN}). 
We subsequently apply Mikkola's method a second time to solve the `quasi-classical'
2PN-accurate KE, namely  $\tilde{l} = u_{\rm 2PN} - e_t\,\sin(u_{\rm 2PN})$, where 
$\tilde{l} = l - l_2\,(u, x, e_t)$. (More details on the implementation 
at the 2PN order can be found in Ref.~\cite{TG07}.)
This allows us to relate $u$ to $l$ (or time) at the 2PN level and therefore to describe 
the temporal evolution of the dynamical variables due to the conservative 2PN-accurate 
orbital dynamics. The use of the above semi-analytic approach 
 ensures that the orbital time scale variations 
are included in a computationally inexpensive way while trying to 
obtain $h_{+,\times} \big|_{\rm Q}(t)$.

 Now let us describe how we implement the secular evolution of the two orbital 
 elements and two angular variables that appear in the PN-accurate Keplerian description.
The secular evolution of the orbital elements, namely $x$ and $e_t$, arises 
due to the effects of GW emission. In contrast, the differential equations for 
$l$ and $\lambda$ are due to the conservative 2PN-accurate orbital dynamics.
Additionally, the use of the above four differential equations ensures that our 
time-domain approximant reduces to the 2PN order {\texttt{TaylorT4}} approximant 
in the circular limit.
We begin by explaining the procedure to compute the differential equations for $x$ and $e_t$.  
These two equations  require two crucial inputs, and 
the first input involves 2PN-accurate expressions for  
$e_t^2$ and $\omega$ in terms of ${\cal E}$ and ${\cal J}$ 
in harmonic gauge, extractable from Ref.~\cite{MGS}. 
The 2PN-accurate expressions for the orbital averaged far-zone energy
and angular momentum fluxes, computed in Refs.~\cite{GI97,ABIS},
form the second input.
We employ the energy and angular momentum balance arguments to
obtain differential equations for the secular evolution of $x$ and $e_t$
after taking the time derivatives of the 2PN-accurate expressions for 
$x$ and $e_t^2$, expressed in terms of ${\cal E}$ and ${\cal J}$.
The resulting differential equations for the secular evolution of 
$x$ and $e_t$ may be displayed as 
\begin{subequations}
\label{Eq:evoleq}
\begin{align}
\frac{d  x}{d t}                  \label{Eq:evoleqa}
      &= \eta \, \frac{c^3}{G\,m} \, x^5 \,
	\biggl \{	\frac{192 + 584\,e_t^2 + 74\,e_t^4}{15\,(1 - e_t^2)^{7/2}} \nonumber \\
		&+	\dot{x}^{\rm 1PN}(\eta,\,e_t)\,x
		+	\dot{x}^{\rm 1.5PN}(e_t)\,x^{3/2} \nonumber \\
		&+	\dot{x}^{\rm 2PN}(\eta,\,e_t)\,x^2
	\biggr\}\,, \\      
\frac{d  e_t}{d t}                    \label{Eq:evoleqb}
      &= - \eta \, e_t \, \frac{c^3}{G\,m} \, x^4 \,
	\biggl \{	\frac{304 + 121\,e_t^2}{15\,(1 - e_t^2)^{5/2}} \nonumber \\
		&+	\dot{e}_t^{\rm 1PN}(\eta,\,e_t)\,x 
		+	\dot{e}_t^{\rm 1.5PN}(e_t)\,x^{3/2} \nonumber \\
		&+	\dot{e}_t^{\rm 2PN}(\eta,\,e_t)\,x^2
	\biggr\}\,, 
\end{align}
\end{subequations}
where the explicit expressions for various PN contributions are listed as Eqs.~(\ref{Eq:dxetl_B}) in
the appendix. These equations are consistent with their  equivalent 
versions in Ref.~\cite{ABIS}.
 
  In our approach, we have adapted a computationally efficient way to incorporate the (relative) 1.5PN 
  corrections to $\dot x$ and $\dot{e}_t$.
These  contributions
are due to the dominant order tail effects
that 
 arise from the non-linear interactions between the multipole 
moments of the GW radiation field and the mass monopole of the source. The
tail contributions are non-local in time and therefore 
 hereditary in nature.
 Following  Refs. \cite{BS93,RS97}, we write the orbital-averaged
far-zone energy and angular momentum fluxes as 
\begin{subequations}
 \label{Eq:heredfluxes}
\begin{align}
 {\langle \mathcal{F} \rangle}_{\rm hered} &= \frac{32}{5}\,\frac{c^5}{G}\,{\eta}^2\,x^5\,
\bigl[4\,\pi\,x^{3/2}\,\varphi(e_t)\bigr]\,, \\
 {\langle \mathcal{G} \rangle}_{\rm hered} &= \frac{32}{5}\,c^2\,{\eta}^2\,m\,x^{7/2}\,
\bigl[4\,\pi\,x^{3/2}\,\tilde{\varphi}(e_t)\bigr]\,,
\end{align}
\end{subequations}
where $\varphi(e_t)$ and $\tilde{\varphi}(e_t)$ define certain eccentricity 
enhancement functions. These functions are usually given 
in terms of infinite sums of Bessel functions  $J_n(n e_t)$ and their   derivatives w.r.t $(n\,e_t)$.
%${J'}_n(n e_t)$. 
The presence of infinite sums of such special functions  implies that the  
 numerical evaluation of the eccentricity  enhancement functions can be computationally 
expensive. In this paper,  we  implement the $e_t$  enhancement functions 
with the help of the following 
 rational functions 
of $e_t$:
\begin{widetext}
\begin{subequations}
\label{Eq:enhfunc}
\begin{align}
 \varphi(e_t) &= \left. \biggl(1 + 7.260831042\,e_t^2 + 5.844370473\,e_t^4 + 0.8452020270\,e_t^6 
 + 0.07580633432\,e_t^8 + 0.002034045037\,e_t^{10}\biggr) \middle/ \right. \nonumber \\
 &\biggl(1 - 4.900627291\,e_t^2 + 9.512155497\,e_t^4 - 9.051368575\,e_t^6 
 + 4.096465525\,e_t^8 - 0.5933309609\,e_t^{10} \nonumber \\
 &- 0.05427399445\,e_t^{12} - 0.009020225634\,e_t^{14}\biggr) \,, \\
 \tilde{\varphi}(e_t) &= \left. \biggl(1 + 1.893242666\,e_t^2 - 2.708117333\,e_t^4 
 + 0.6192474531\,e_t^6 + 0.05008474620\,e_t^8 - 0.01059040781\,e_t^{10}\biggr) 
 \middle/ \right. \nonumber \\
 &\biggl(1 - 4.638007334\,e_t^2 + 8.716680569\,e_t^4 - 8.451197591\,e_t^6 + 4.435922348\,e_t^8
 - 1.199023304\,e_t^{10} \nonumber \\
 &+ 0.1398678608\,e_t^{12} - 0.004254544193\,e_t^{14}\biggr)\,.
\end{align}
\end{subequations}
\end{widetext}
The coefficients of the above two rational functions are obtained from the Taylor expanded 
versions of $\varphi(e_t)$ and $\tilde{\varphi}(e_t)$ in the small $e_t$ limit.
The procedure to construct such rational functions is 
 explained in Sec.~8.3 of Ref.~\cite{BO99_book}.
 Invoking the terminology of Ref.~\cite{BO99_book},
 we may refer to the above $\varphi(e_t)$ and $\tilde{\varphi}(e_t)$
 expressions as Pade approximants $P^{5}_{7}( e_t^2)$.
We have verified that the numerical estimates of $\varphi(e_t)$ and $\tilde{\varphi}(e_t)$ 
that are based on our Eqs.~(\ref{Eq:enhfunc}) match accurately with those listed in Tables I and II
in Ref.~\cite{ABIS} for $e_t \leq 0.9$.
We observe that
Ref.~\cite{ABIS} obtained these 
$e_t$  enhancement 
functions numerically through a Fourier analysis of quasi-Keplerian motion. 
This gives us confidence in applying the above rational functions to compute 
the tail contributions to $\dot x$ and $\dot{e}_t$.
 With the help of the above two rational functions, it is fairly straightforward to obtain
the 1.5PN-accurate tail contributions to the differential equations for $x$ and $e_t$ as
\begin{subequations}
 \label{Eq:heredevoleq}
\begin{align}
 \dot{x}^{\rm 1.5PN}(e_t) &= \frac{64}{5}\,\biggl[4\,\pi\,\varphi(e_t)\biggr]\,, \\
 \dot{e}_t^{\rm 1.5PN}(e_t) &= \frac{32}{5}\,\biggl[\frac{985}{48}\,\pi\,\varphi_e(e_t)\biggr]\,,
\end{align}
\end{subequations}
where
\begin{align}
 \label{Eq:phieofe}
 \varphi_e(e_t) &= \frac{192}{985}\,\frac{\sqrt{1 - e_t^2}}{e_t^2}\,
 \biggl[\sqrt{1 - e_t^2}\,\varphi(e_t) - \tilde{\varphi}(e_t)\biggr]\,.
\end{align}

 Let us now explain why we need two additional differential equations to specify 
 $l$ and $\lambda$ evolutions.
 Recall that we require to specify the values of $x,e_t, l$ and $\lambda$ at each instant to
 obtain the temporally evolving $h_{+,\times} \big|_{\rm Q}(t)$ for 
 compact binaries that are specified by certain values of 
$x,e_t, c_l$ and $c_{\lambda}$ at the initial epoch.
Therefore, it is very convenient to provide differential equations for describing the temporal evolution of $l$ and $\lambda$ 
for binaries inspiraling along PN-accurate eccentric orbits.
Additionally, the differential equation for $\lambda$ ensures that in the circular limit our eccentric approximant 
goes to the {\texttt{TaylorT4}} approximant, as $\phi = \lambda$ in this limit.
The differential equations for $l$ and $\lambda$  are given by
\begin{subequations}
\label{Eq:evol_llambda}
\begin{align}
        \frac{d  l}{d t} &= n 
      = x^{3/2} \, \frac{c^3}{G\,m}
	\biggl \{	1
		+	\dot{l}^{\rm 1PN}(e_t)\,x \nonumber \\
		&+	\dot{l}^{\rm 2PN}(\eta,\,e_t)\,x^2
	\biggr\}\,, \\
\frac{d  \lambda}{d t} &= \omega 
      = x^{3/2} \, \frac{c^3}{G\,m}\,,
\end{align}
\end{subequations}
and the PN contributions are once again listed in the appendix.
It is not very difficult to deduce that the equation for $\dot l$ arises 
from the PN-accurate relation connecting $n$ and $\omega$ (see Ref.~\cite{ABIS}).
The differential equation for $\lambda$ is due to the fact that the orbital averaged differential 
equation for $\phi$, namely $\left\langle d \phi/dt \right\rangle$, is identical to $d \lambda /dt = \omega$.
With the listing of the above two equations, we have all the ingredients to obtain
 $h_{+,\times} \big|_{\rm Q}(t)$ for compact binaries inspiraling under the 
 influence of 2PN-accurate GW emission while moving along 2PN-accurate 
 eccentric orbits.
 
 Observe that the $e_t$ contributions are treated in an exact manner 
in all the instantaneous contributions to the  differential equations for $x,e_t$ and $l$
that appear at the Newtonian, 1PN and 2PN orders.
The use of rational functions at 1.5PN order ensures that we also have closed form expressions
to  evolve $x$ and $e_t$.
Note that it is the use of rational functions, analytic expressions for 
the dynamical variables and Mikkola's method that make our approach 
numerically accurate and possibly 
computationally  less expensive than the $x$-model at this PN order.
However, further investigations
 will be required to quantify this observation.

Let us now briefly explain 
the time-domain 
$x$-model, proposed in Ref.~\cite{Hinder09}, to compute 
 PN-accurate waveforms for eccentric inspirals.
 This model also invokes $x$ and $e_t$ to characterize the binary orbit.
Therefore, the effects of GW emission on the usual dynamical variables
%namely $r, \dot r, \dot \phi$ and $\phi$,
are included by solving 2PN-accurate 
 differential equations for $x$ and $e_t$. 
Certain numerical fits are employed  to model the 1.5PN order tail contributions
 to $\dot x$ and $\dot{e}_t$, as detailed in Ref.~\cite{Hinder09}.
 However, the conservative dynamics is 3PN-accurate and the associated parametric 
expressions for the dynamical variables can be quite lengthy.
The $x$-model employs PN-accurate expressions for $ r $ and $\dot \phi$ 
in terms of $x,e_t$ and $u$; hence the $x$-model also requires solving the PN-accurate Kepler equation to 
model the conservative temporal evolution of these dynamical variables.
In contrast to our approach,
Ref.~\cite{Hinder09} numerically differentiates and integrates 
the parametric expressions for $ r $ and $\dot \phi$ to obtain 
values of $\dot r$ and $ \phi$ at each time step.
This was pursued due to the lengthy nature of the these dynamical variables.
In our opinion, the use of numerical integration and differentiation at every time step to obtain 
 $h_{+,\times}|_{Q}(t)$ may 
make  the $x$-model  
 computationally more demanding than our present approach.
 As noted earlier, further investigations involving our approach at the fully 3PN 
 order for binaries with arbitrary $\eta$ values should be pursued to clarify the above observation. 
 
 To operationalize our prescription, we choose certain $(x,e_t, c_l, c_{\lambda})$
 values at an initial epoch to specify our $(m,\eta, \iota)$ binary.
 The use of 2PN-accurate KE results in the corresponding value
 for  $h_{+,\times}|_{Q}(t)$ at that initial epoch.
 We numerically solve  simultaneously the four differential equations for $x,e_t, l$ and 
 $\lambda$ to obtain values of these variables at $t_0+ \Delta t$.
 The use of KE at that step results in unique values for  
  $h_{+,\times}|_{Q}(t)$ at that instant. We repeat these steps till $x$ reaches its cut-off value 
  of $\sim 0.1667$.
  This value arises as we terminate the orbital evolution 
  when the orbital separation reaches  the value associated with 
  the last stable orbit (LSO) for a test particle in a Schwarzschild space-time, namely  
   $r_{\rm LSO}=6\, G \,m/c^2$.
 This leads to the above $m$ independent value for $x$.
 Therefore, we clearly do not include eccentricity effects on our termination value for $x$.
Additionally, we do not consider the possibility that eccentric orbits near 
the LSO may not obey a Keplerian-type parametric solution as noted 
in Ref.~\cite{DGI}.
We plan to investigate these subtle issues in another paper.
In what follows, we display GW polarizations states of our approach
and probe preliminary data analysis implications.

\subsection{ Eccentric TaylorT4 approximant: facets and implications }

\begin{figure*}[htp]
\begin{center}
\includegraphics[width=0.8\textwidth, angle=0]{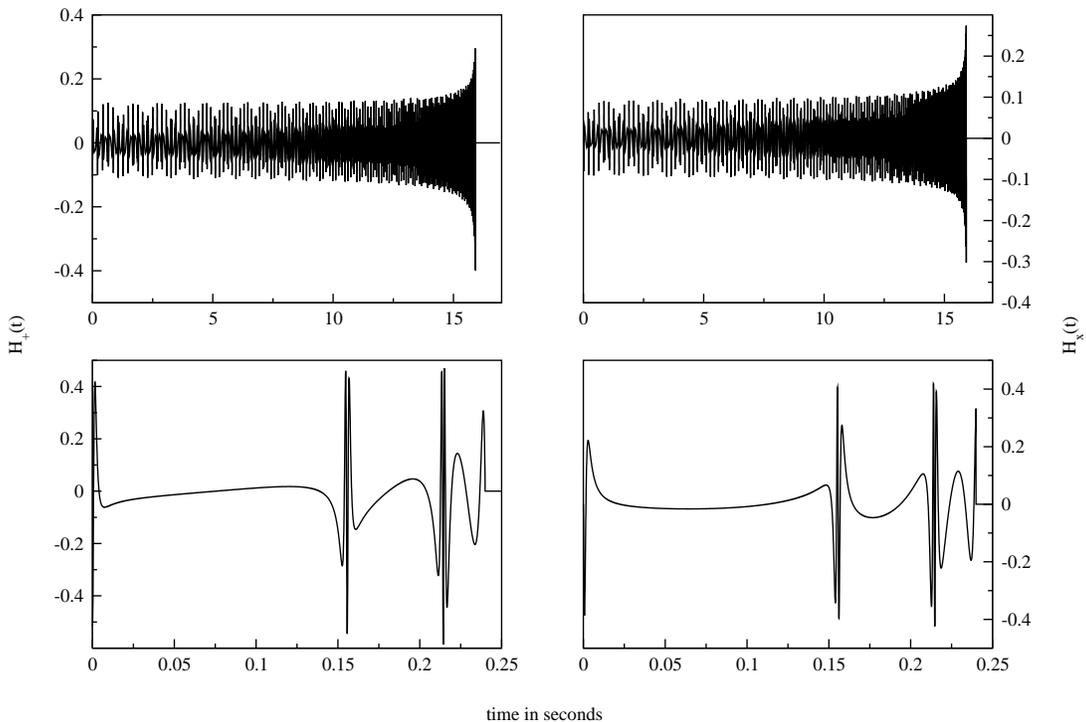}\\[0.1cm]
\end{center}
\caption{\label{FIG:waveformplots}Plots that display temporally evolving
scaled GW polarization states $H_{+,\times}\big|_{\rm Q}(t)$ for 
stellar-mass BH-BH binaries with $m_1 = m_2 = 10\, M_{\odot}$ and 
for two $e_0$ values.
The plots in the upper and lower panels consider binaries with $e_0 = 0.45$ and $0.85$, respectively.
We observe chirping GW signals modulated by the influence of periastron advance in 
the upper panel plots.
The lower panel plots depict the `repeated burst' nature of highly eccentric 
inspirals.
}
\end{figure*}

We are now in a position to numerically implement $h_{+,\times} \big|_{\rm Q}(t)$ 
that model GWs from 
non-spinning compact binaries inspiraling along 2PN-accurate eccentric orbits under the influence of GW emission at 2PN order.
In Fig. \ref{FIG:waveformplots}, we display temporal plots for the scaled 
$h_{+,\times}\big|_{\rm Q}(t)$ for $m_1=m_2 = 10\,M_{\odot}$ BH-BH binaries 
with moderate and high initial eccentricities in the aLIGO frequency window, namely $e_0=0.45$ and $e_0 = 0.85$, respectively.
In these plots, $H_{+,\times}(t)$ stand for $h_{+,\times}\big|_{\rm Q}(t)$ scaled by $G\, m \eta /(c^2 D_L)$.
The upper row of plots shows $H_{+,\times}(t)$ for binaries with $e_0=0.45$ while the lower row of plots depicts 
the GW polarization states when $e_0 = 0.85$, with $\iota=\pi/3$ in both cases.
The plots for the case of moderate initial eccentricity clearly show the chirping of GW signals that are modulated by the 
advance of periastron. For high initial eccentricity, we observe that the GW signal consists of a series of 
repeated bursts; these bursts of GWs arise from the successive periastron passages.
Note that the time intervals between successive peaks of GWs decrease substantially.
This is due to the $e_t$ induced shortening of the GW radiation reaction time scale.
As a result, the signal duration of the $e_0 = 0.85$ inspiral in the aLIGO frequency 
window is substantially shorter compared to the $e_0=0.45$ inspiral.

Strictly speaking, our approach to 
model aLIGO inspirals with high $e_0$ values can be problematic.
This is because of the possibility that the Keplerian-type 
parametric solution may not be appropriate to model 
such highly eccentric and relativistic orbits as noted in Ref.~\cite{DGI}.
It will be desirable to adapt and extend the 
 Effective-One-Body (EOB) formalism for general orbits, detailed in  Ref.~\cite{BD12}, to model GWs from such binaries.

 \begin{figure*}[htp]
\begin{center}
\includegraphics[width=0.8\textwidth, angle=0]{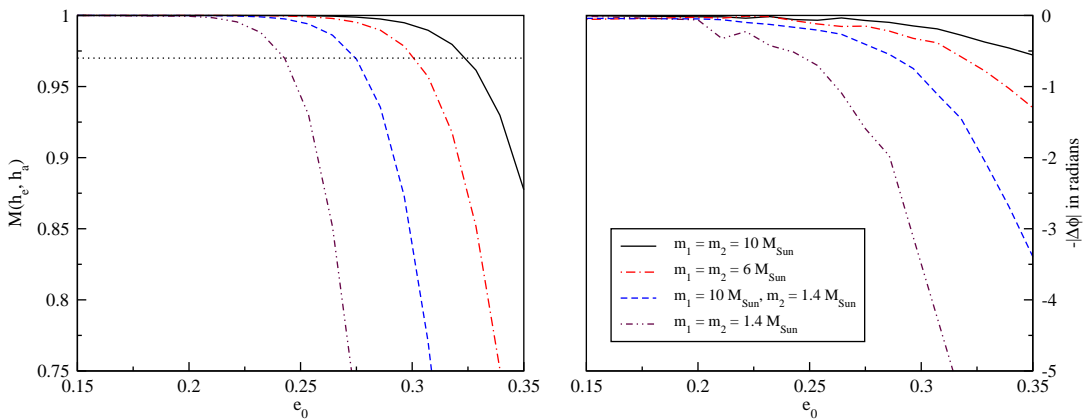}\\[0.1cm]
\end{center}
\caption{\label{FIG:matchetvsTe8}(Color Online) Plots that probe the total-mass dependence of our ${\cal M}(h_e, h_a)$ and the related $\lvert \Delta \phi (h_e,h_a) \rvert$ estimates
as functions of $e_0$ 
for a number of aLIGO relevant stellar-mass compact binaries.
The $h_e$ inspiral waveforms are based on our eccentric {\texttt{TaylorT4}} approximant while 
$h_a$ waveforms arise from the Te8 approximant, as detailed in the text.
The critical $e_0$ values are higher for higher mass binaries due to their shorter inspiral lifespan
in the aLIGO frequency window.
A clear correlation exists between the drop in  the ${\cal M}( h_e, h_a) $ values and their associated
$\Delta \phi (h_e,h_a)$ values.
}
\end{figure*}

\begin{figure*}[htp]
\begin{center}
\includegraphics[width=0.8\textwidth, angle=0, clip=true]{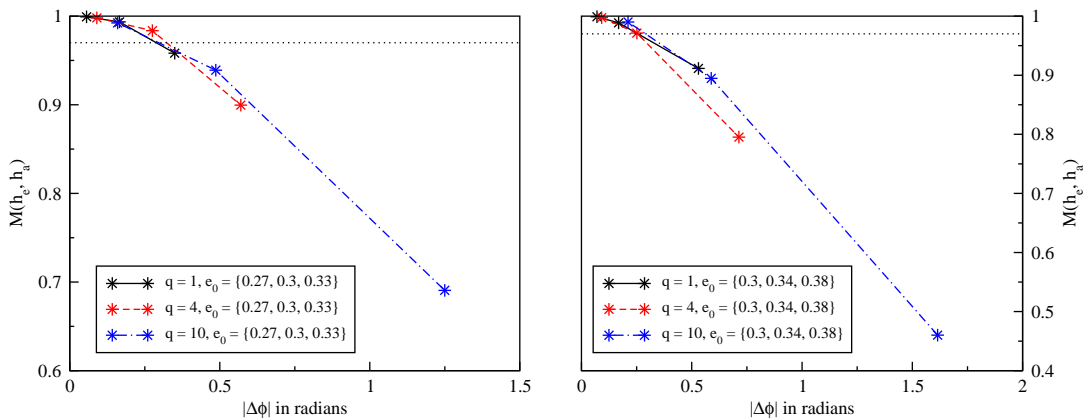}\\[0.1cm]
\end{center}
\caption{\label{FIG:matchetvsTe8_2}(Color Online) Plots that probe the dependence of our $ {\cal M}( h_e, h_a)$ estimates on the mass ratio $q$.
We display a set of $\{ {\cal M}( h_e, h_a), \lvert \Delta \phi (h_e,h_a) \rvert \} $ 
 values for BH-BH binaries with different $q$ and $e_0$ values.
 The left panel considers binaries with  $m = 20\, M_{\odot}$, while the right panel contains 
 data points for binaries with $m = 40\, M_{\odot}$.
Sharper drops in the match estimates are clearly visible for binaries with larger $q$.
 The neglected orbital eccentricity contributions in the $h_a$ waveforms 
 force them to dephase strongly from
 their $h_e$ counterparts during the comparatively longer aLIGO evolution window for the binaries with larger $q$.
}
\end{figure*}

An important aspect of our approach is its ability to treat the eccentricity contributions in an exact manner 
while modeling the time-domain inspiral waveforms.
In this context, the `exact in $e_t$'  feature of our approach refers to 
the fact that the $e_t$ contributions are incorporated in a non-perturbative manner.
This is mainly due to the use of a PN-accurate Keplerian-type parametric solution while tackling
the conservative dynamics and the use of rational functions while incorporating the tail contributions 
into the PN-accurate $\dot x$ and $\dot{e}_t$ expressions.
Note that the instantaneous contributions to  
the differential equations for $x$ and $e_t$, appearing at
Newtonian, 1PN and 2PN order, 
treat $e_t$ in an exact manner due to the use of PN-accurate Keplerian-type parametric solution
for the orbital averaging \cite{GI97}.
This results in closed form expressions 
for the instantaneous contributions to $\dot x$ and $\dot{e}_t$.
Strictly speaking,  our 2PN-accurate approximant treats $e_t$ contributions in an exact manner, provided 
$e_0 < 0.9$, due to the use of rational functions at 1.5PN order as noted earlier.
This feature allows us to employ our approximant to probe 
the GW data analysis implications of using inspiral templates where orbital eccentricity effects are treated in an approximate manner, especially while incorporating the effects of GW emission.
For this purpose, we construct a second time-domain inspiral family, namely a ${\rm Te8}$ approximant, by Taylor expanding 
the differential equations for $x$ and $e_t$, given by Eqs.~(\ref{Eq:evoleqa}) and (\ref{Eq:evoleqb}), 
in the small $e_t$ limit while keeping $e_t$ contributions accurate up to ${\cal O}(e_t^8)$.
This ${\rm Te8}$ approximant is motivated by our observations that one will be forced to 
 treat $e_t$ in an approximate manner if one wishes to incorporate 
 effects of orbital eccentricity into other circular time domain approximants like
 {\texttt{TaylorT1}}, {\texttt{T2}}, and {\texttt{T3}}
  in a straightforward manner.
  We are 
restricting $e_t$ contributions to the eighth order to be
consistent with the order of 
 eccentricity corrections available in the PC 
 and EPC prescriptions \cite{YABW09,H14}.
The match (${\cal M}$) estimates,
%detailed in Refs.~\cite{BO95,DIS98,ajith2011}, 
detailed in Refs.~\cite{BO95,DIS98},
are invoked to compare 
the $e_t$ exact and $e_t$ truncated waveform families.
%http://adsabs.harvard.edu/abs/1996PhRvD..53.6749O
In particular, we explore the faithfulness of the $e_t$ truncated waveform family. 
Faithfulness requires that the associated match $({\cal M})$ values are greater than $0.97$.
A few comments are in order before we proceed with the match computations. 
It should be clear that 
the ${\cal M}$ estimates between the above two
approximants probe only the
consequence of truncating $e_t$ contributions while constructing time-domain inspiral waveform families.
At present, we do not explicitly pursue match computations between our 2PN order
eccentric extension of the
{\texttt{TaylorT4}} model and the various 
%eccentricity-approximate
 Fourier-domain models,available in Refs.~\cite{YABW09,H14} 
 and in the previous section.
This is mainly due to the model-dependent systematic mismatch that occurs while
comparing even the time-domain and
frequency-domain quasi-circular inspiral templates~\cite{BIOPS}. Therefore, it is
reasonable to expect similar systematic effects
when considering eccentric inspiral templates in the time and frequency domain.
However, we would like to emphasize that in the $e_0 \rightarrow 0$ limit our fully
analytic frequency-domain waveforms $\tilde h(f)$
reduce to the {\texttt{TaylorF2}} model at 2PN order exactly.
Finally, in the following (see Figs. \ref{FIG:matchetvsTe8} and
\ref{FIG:matchetvsTe8_2}) $h_e$ stands for our eccentric extension
of the {\texttt{TaylorT4}} approximant where
$e_t$ contributions are treated in an exact manner while $h_a$  stands for time domain waveform families based on the truncated ${\rm Te8}$ approximant.

The ${\cal M}(h_e, h_a)$ computations that we pursue here require us to define  
a certain overlap function between 
our time domain $h_e $ and $ h_a$ inspiral waveform families where $h_e$ and $h_a$ refer to the respective cross polarization states of the GW.
The overlap integral 
$\mathcal{O}(h_e, h_a)$ is defined as 
\begin{align}
\label{Eq:overlap}
 \mathcal{O}(h_e, h_a) &= 
 %\left\langle \hat h_e, \hat h_a  \right\rangle  =
\frac{\langle h_e|h_a \rangle}{\sqrt{\langle h_e|h_e \rangle \, \langle h_a|h_a\rangle}} \,.
\end{align}
Clearly, the overlap integral requires a certain normalized inner product involving the
$h_e(t)$ and $h_a(t)$ families. This is given by
\begin{align}
\label{Eq:innerproduct}
\langle  h_e |  h_a \rangle &= 4\, {\rm Re}\,  \int_{f_{\rm low}}^{f_{\rm cut}} \, 
\frac{\tilde h_e^*(f)\, \tilde h_a(f)}{S_{\rm h}(f)} df \,.
\end{align}
The symbols $\tilde h_e(f)$ and $\tilde h_a(f)$ stand for  the Fourier transforms of the $h_e(t)$ and
$h_a(t)$ inspiral waveforms 
while $S_{\rm h}(f)$ denotes
the one-sided power spectral density of the detector noise. We have used the 
zero-detuned, high power (ZDHP) noise configuration of aLIGO, provided in Ref.~\cite{LIGO_2010}, for the present
${\cal M}(h_e, h_a)$ computations.
For these match estimates, we let $ f_{\rm low}$ be $10\,$Hz, corresponding to the lower cut-off frequency of aLIGO,
while the upper cut-off frequency, as noted earlier, is chosen to be $f_{\rm LSO}=c^3/(G\, m\, \pi\, 6^{3/2})$.
The match ${\cal M}(h_e,h_a)$ is obtained by maximizing the above overlap over 
certain kinematical variables of $h_e$ such that 
\begin{align}
\label{Eq:match}
{\cal M}(h_e, h_a) &= \max_{t_0, \phi_0}\, \mathcal{O}(h_e, h_a)\,,
\end{align}
where $t_0$ and $\phi_0$ are the detector  
arrival time  and the associated  phase $\phi_0$ of the $h_e$ template.
The maximization over $t_0$ is performed with the help of the FFT algorithm, 
while we apply two orthogonal templates to maximize over $\phi_0$ \cite{DIS98}.
In our match estimates,  all other  parameters pertaining to both $h_e$ 
and $h_a$ waveform families are treated to be identical.
Let us emphasize that  ${\cal M}(h_e,h_a)>0.97 $ implies that the 
approximate  ${\rm Te8}$  waveform family will recover our time-domain approximant,
constructed to be an 
eccentric extension of the 2PN-accurate {\texttt{TaylorT4}} approximant, in a `faithful' manner.

   In Fig.~\ref{FIG:matchetvsTe8}, we plot 
the ${\cal M}(h_e, h_a)$ estimates and the absolute values of the related accumulated 
phase differences, namely $\lvert \Delta \phi \rvert$, as functions of $e_0$ for typical aLIGO relevant compact binaries containing NSs and BHs.
The plots indicate that ${\cal M}( h_e, h_a) $ values drop below $0.97$ 
when $\lvert \Delta \phi (h_e,h_a)  \rvert $ values are $\sim 0.5$ radians.
A direct correlation between the drop in  ${\cal M}( h_e, h_a) $ values and their 
$-\lvert \Delta \phi (h_e,h_a)  \rvert$ values is also observed.
The critical values $e_0^{c}$ for initial eccentricity, above which the ${\cal M} \leq 0.97$, lie roughly
in the $0.25-0.35$ range for our aLIGO binaries.
These $e_0^c$ values clearly depend on the total mass $m$, 
and we observe that the $\{h_e, h_a \}$ templates with higher total mass dephase at higher $e_0$ values 
compared to their lower mass counterparts.
A possible explanation is that the lower mass binaries last longer in the aLIGO frequency window
which provides more time for even small differences between the two approximants to grow.
For higher mass binaries,  comparable changes in $\lvert \Delta \phi (h_e,h_a) \rvert$ values
occur for higher $e_0$ values due to their shorter lifespan in the aLIGO band.
 This is also clearly evident while comparing the plots for the NS-NS and BH-BH binaries.
The observed direct correlation between the drop in  ${\cal M}( h_e, h_a) $ values and their 
$-\lvert \Delta \phi (h_e,h_a)  \rvert$ values has direct implications for 
our $\tilde h(f)$ computations, detailed in Sec.~\ref{sec:eTF2}, and the numbers listed in Table~\ref{table:1}.
This correlation implies that match computations between our 2PN order $\tilde h(f)$
and its EPC counterpart should yield ${\cal M}$ values substantially lower than $0.97$ for 
configurations with $e_0\sim 0.1$.
Therefore, including the effects of orbital eccentricity evolution into $\tilde h(f)$ in a PN-accurate manner is also important from the perspective of match computations.

 Let us move on to probe the influence of the mass ratio $q= m_1/m_2$ on our ${\cal M}(h_e, h_a)$ estimates.
 In Fig.~\ref{FIG:matchetvsTe8_2}, we display a set of $\{  {\cal M}( h_e, h_a),\lvert \Delta \phi (h_e,h_a) \rvert\}$ 
 values for BH-BH binaries with different $q$ and $e_0$ values.
 The plots in the left and right panels consider binaries with total mass $m = 20\, M_{\odot}$ and $ 40\, M_{\odot}$, respectively. The $e_0$ values have been chosen so that the maximum  $|\Delta \phi (h_e,h_a)| $ value will not exceed
 two radians. We observe a sharp drop in ${\cal M}( h_e, h_a)$ values for binaries with larger $q$.
 This may also be related to the fact that such binaries last longer in the aLIGO frequency window
 compared to their counterparts with lower $q$ value. 
 The neglected orbital eccentricity contributions force the $e_t$ truncated $h_a$ templates to dephase strongly from
 their $e_t$ exact $h_e$ counterparts during the comparatively longer aLIGO evolution window for the binaries with larger $q$.
 This dephasing provides a natural explanation for the sharp drop in the ${\cal M}( h_e, h_a)$  
 estimates for such binaries, as is evident from the plots for binary configurations with smaller $m$ and larger $q$.

\begin{figure*}[htp]
\begin{center}
\includegraphics[width=0.8\textwidth, angle=0]{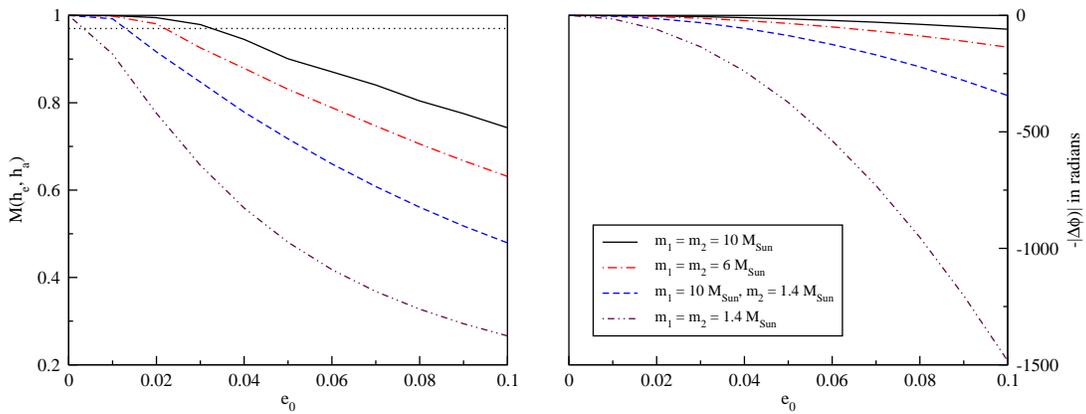}\\[0.1cm]
\end{center}
\caption{\label{FIG:matchetvscir}(Color Online) Plots that probe the faithfulness of the quasi-circular 2PN {\texttt{TaylorT4}} approximant with respect to its 
eccentric extension. The ${\cal M}( h_e, h_a)$ estimates and the related $\lvert \Delta \phi (h_e,h_a) \rvert$  
as functions of $e_0$ are considered for typical aLIGO binaries.
Clearly, the quasi-circular templates are faithful to our eccentric {\texttt{TaylorT4}} approximant only
for binaries with tiny residual eccentricities like $10^{-2}$ or less. Hence, such templates can 
capture an eccentric GW signal only if the
binary has a tiny residual eccentricity when the system enters the aLIGO band.
}
\end{figure*}

 % Finally in Fig.~\ref{FIG:matchetvscir}, 
  We probe the ability of the 2PN-accurate quasi-circular {\texttt{TaylorT4}} approximant to faithfully capture its eccentric extension in Fig.~\ref{FIG:matchetvscir}.
 We plot  ${\cal M}( h_e, h_a)$  estimates and the related $\lvert \Delta \phi (h_e,h_a)  \rvert$ 
 as functions of $e_0$, 
 where $h_a$ now stands for inspiral templates based on  
 the quasi-circular 2PN-accurate {\texttt{TaylorT4}} approximant.
 The quasi-circular templates can faithfully capture an eccentric GW signal only if the
binary has a tiny residual eccentricity when the system enters the aLIGO band.
The $e_0^c$ value above which the match estimates fall below $0.97$ is again $m$ dependent.
This critical initial eccentricity is $\sim 0.04$ for  BH-BH binaries with $m = 20 M_\odot$, while 
$e_0^c \sim 0.005$ for NS-NS binaries. For low mass binaries, the neglected eccentricity contributions force the $h_a$ templates to dephase strongly from their $h_e$ counterparts, due to the comparatively longer inspiral time of such binaries in the aLIGO window. This provides the expected explanation for the $m$ dependency of the critical $e_0^{c}$ values. 
These observations are consistent with the fitting factor calculations that probed
the ability of the quasi-circular {\texttt{TaylorT4}} approximant to detect GWs from eccentric binaries constructed with the help of the $x$-model \cite{HB2013}.

Finally, we invoke the above match plots 
and the {\it 2PN analytic} entries of Table.~\ref{table:1}
to provide a non-rigorous justification for truncating 
our frequency-domain approximant to ${\cal O}(e_0)^6$.
The match plots of the present section reveal that 
our approximate time domain 
waveform family with eccentricity contributions accurate up to
${\cal O}(e_0)^8$ is faithful to our {\it exact in $e_t$} time-domain approximant only 
for initial eccentricities $e_0 \leq 0.25-0.35$,
depending on the total mass of the binary. 
We  have also checked that 
the fractional differences in ${\cal N}$,
associated with the {\it 2PN analytic} entries, 
are also fairly constant while considering initial eccentricities up to $0.25$.
This  suggests that
including eccentricity contributions accurate to ${\cal O}(e_0)^6$ 
is sufficient from the point of 
view of accumulated number of GW cycles ${\cal N}$.
Therefore, it may be reasonable to expect that {\it approximate in 
eccentricity} frequency-domain inspiral families are also faithful to our $e_t$ exact time-domain approximant only up 
to such initial $e_0$ values.
Additionally, we observe that the {\it 2PN analytic} ${\cal N}$  estimates do not change 
substantially when we drop the ${\cal O}(e_0)^6$ contributions to $\phi (x,e_0,F_0)$ 
for binaries with initial eccentricities up to $0.25$.
These considerations, in our opinion, provide reasonable justification  
for restricting 
the initial 
eccentricity corrections up to ${\cal O}(e_0)^6$ at each PN order in our frequency-domain approximant and therefore to let the harmonic index $j$ vary up to $8$ in Eq.~(\ref{10}).
Clearly, match estimates involving frequency and time domain waveform families that are accurate to 3PN 
order will be desirable to check the validity of these statements.
We expect that this extension should also clarify the need to 
go beyond the ${\cal O}(e_0)^6$ corrections from the point of 
view of the lengthy expressions for $\Psi_j$ and $e_t (\omega)$.

\section{Conclusions}            
\label{sec:conclusions}

In this paper,  we computed 
a fully analytic frequency-domain 
inspiral waveform with Newtonian amplitude 
and 2PN order  Fourier phase while incorporating 
eccentricity effects up to sixth order at each PN order.
This is achieved by extending the post-circular scheme of Ref.~\cite{YABW09}
by incorporating the effects of 
PN-accurate orbital eccentricity evolution.
With the help of the accumulated number of GW cycles 
in a certain $x_{\rm low}-x_{\rm high}$ window,
suitable for the advanced GW detectors,
 we showed the importance of incorporating 
 eccentric contributions to
the Fourier phase in a PN consistent manner.
We also presented a prescription 
 to incorporate orbital eccentricity into the quasi-circular 
time domain {\texttt{TaylorT4}} approximant at 2PN order. 
This involved employing 
rational functions in orbital eccentricity  to  
implement the 1.5PN order tail contributions to the far-zone
fluxes and a modified version of Mikkola's method 
to solve the PN-accurate Kepler equation.
Our approach contains 
  closed form PN-accurate differential
equations for evolving PN-accurate eccentric orbits while 
treating eccentricity effects in an exact manner.
We point out that our  
time domain eccentric approximant  should be  accurate and 
efficient to handle initial orbital eccentricities $\leq 0.9$.
With the help of match estimates, 
preliminary GW data analysis implications are 
probed.
We note in passing that the above prescriptions for 
eccentric inspiral templates have been implemented 
in the LSC Algorithm Library of the LIGO Scientific Collaboration.

A number of extensions are possible and some of these are being actively pursued.
 Indeed, it is possible to extend the PN-accuracy of both approximants to the next PN order.
 For the frequency-domain waveforms, we require 3PN-accurate expressions for 
 $\dot {\omega}$ and $\dot{e}_t$, available in Ref.~\cite{ABIS}.
At present, efforts are on-going to extend the analysis of 
Ref.~\cite{Favata2014} by incorporating 
$e_0^2$ contributions to the SPA phase at 3PN order \cite{Arun15}.
 It should also be possible to include the effect of periastron advance by adapting and 
 extending the arguments present in Sec.~VI of Ref.~\cite{YABW09}.
 Another direction of investigation will be to incorporate PN order amplitude contributions to 
 our $\tilde h(f)$ with the help of Refs.~\cite{TS10,TS11}.
 In comparison, we will require the 3PN-accurate Keplerian-type parametric solution of 
 Ref.~\cite{MGS} and the 3PN-accurate  
 $\dot {\omega}$ and $\dot{e}_t$ expressions  of  Ref.~\cite{ABIS} to extend our 
 $e_t$ exact time-domain approximant to the next PN order.
 This extension should allow us 
to estimate  the comparative accuracies  and efficiencies of the $x$-model
and our fully 3PN-accurate time-domain approximant while 
considering compact binaries with arbitrary (but allowed) $\eta$ and $e_0$ values.
It should be possible to improve Ref.~\cite{GS11} to include the dominant order 
 spin-orbit interactions in our time domain approximant.
 These PN extensions should allow one to pursue detailed comparisons with 
 numerical relativity based eccentric inspirals, thereby extending  the earlier 
 comparison of Ref.~\cite{Hinder09}.
 In this context, it will be interesting to compare our PN-accurate $h(t)$ with the 
 Effective-One-Body (EOB) based eccentric $h(t)$ family.
 This requires adapting the formalism of Ref.~\cite{BD12} to obtain 
 the EOB based $h(t)$ during the inspiral phase.
 
  It will also be of interest to compare our $h(t)$ with the time domain waveforms based on the {\texttt{CBwaves}} software, discussed in Ref.~\cite{Vasuth12},  while considering non-spinning 
  compact binaries in inspiraling eccentric orbits.
  {\texttt{CBwaves}} numerically integrates 3.5PN-accurate equations of motion 
  to incorporate the dynamics of inspiraling eccentric compact binaries into the GW polarization
  states. 
  A similar approach was employed to model GWs from dynamically formed  highly eccentric 
  binaries that can last minutes to days before coalescence \cite{KL12}.
  Clearly, PN extensions of our $h(t)$ will be useful to obtain accurate 
  GW templates for such a `repeated bursts' scenario and to probe its implications.
A possible comparison of our $h(t)$ in the small $\eta$ limit with the
GW strain of  Ref.~\cite{East2013}
should be helpful to mark the $\eta$ range of these two approaches.    
The present approach, capable of modeling highly eccentric inspirals, 
should be interesting to various non-optimal excess power methods
to search for GW bursts \cite{Sergey08,TMP14,CMTLC14}.
A post-Newtonian accurate analytic approach to describe the evolution of $\omega$ and $e_t$ 
should be useful for the seedless clustering approach of Ref.~\cite{CMTLC14}.

\section{Acknowledgements}
We thank K.~G.~Arun, M.~Favata, 
A.~Gupta and E.~A.~Huerta for helpful discussions and suggestions.
Additional thanks to  K.~G.~Arun
for providing us the numerical values for the eccentricity enhancement functions.

\onecolumngrid
\pagebreak

\twocolumngrid
\appendix

\section{  Explicit  2PN order 
$ \Psi_j$, $e_t$ and $\phi$   expressions}   \label{appendixA}

We list below the main and lengthy results of our Sec.~\ref{sec:eTF2}.
Invoking  the convention and symbols of  Sec.~\ref{sec:eTF2}, we write the analytic 
frequency-domain GW strain for eccentric inspirals with Newtonian 
amplitude and 2PN order phase as

\begin{widetext} 
%he Fourier Transform of the response function $h(t)$ is given as
\begin{equation}
\tilde{h}(f) =  \mathcal{\tilde{A}}    {\left(\frac{G m \pi f}{c^3}\right)}^{-7/6}     \sum\limits_{j=1}^{8} \xi_{j}
{\left(\frac{j}{2}\right)}^{2/3}  e^{-i(\pi/4 + \Psi_j)} \,,          
\end{equation}
where $\mathcal{\tilde{A}}$ and $ \xi_j $ are defined as
\begin{subequations}                         
\begin{align}
 \mathcal{\tilde{A}} &= - {\left(\frac{5 \eta \pi}{384}\right)}^{1/2}  \frac{G^2 m^2}{c^5 D_L}  ,  \\             
 \xi_j  &=  \frac{\left(1-e_t^2\right)^{7/4}}{{\left(1+\frac{73}{24}e_t^2+\frac{37}{96}e_t^4\right)}^{1/2}} \alpha_{j} e^{-i \phi_j(f/j)   }  \,.                                                       
\end{align}
\end{subequations}

\end{widetext}

We do not list explicitly  the coefficients $\xi_j$ as polynomials in $e_t$ while incorporating 
$e_t$ contributions up to ${\cal O}(e_t^6)$ as required.
Clearly, it is fairly straightforward to obtain such  $\xi_j$ expressions  from its above definition. 
The explicit expression for the 2PN-order Fourier phase $\Psi_j$ that includes all ${\cal O}(e_0^6)$ contributions is given by

\begin{widetext}
\begin{align}                \label{appendixpsi}
  \Psi_j &\sim 
    j \phi_c - 2\pi f t_c - \frac{3}{128 \eta} \left(  \frac{G m \pi f}{c^{3}}\right)^{-5/3}        \left(\frac{j}{2}\right)^{8/3}           \sum_{n=0}^{4} \mathcal{C}_n x^{n/2},                              
\end{align}

where the coefficients $\mathcal{C}_n$ can be listed as 
\begin{subequations}
\begin{align}
 &\mathcal{C}_0 = 1- \frac{2355}{1462}e_0^{2}\chi^{-19/9} + \left( -  \frac{2608555}{444448}\chi^{-19/9}+ \frac{5222765}{998944}\chi^{-38/9} \right)e_0^{4}  + \left(- \frac{1326481225}{10134144}\chi^{-19/9}     \right.\nonumber\\ 
  &\qquad \left.{} + \frac{173355248095}{455518464}\chi^{-38/9}- \frac{75356125}{3326976}\chi^{-19/3}  \right)e_0^{6},   \nonumber\\ \\  
 &\mathcal{C}_1= 0  ,  \\ \nonumber\\
 &\mathcal{C}_2 = \frac{3715}{756} + \frac{55}{9}\eta + \left\lbrace    \left(-\frac{2045665}{348096}-\frac{128365}{12432}\eta \right)\chi^{-19/9}  +              \left(-\frac{2223905}{491232}+\frac{154645}{17544}\eta \right)\chi^{-25/9}   \right\rbrace e_0^{2}               						\nonumber\\
 &\qquad       + \left\lbrace    \left(-\frac{6797744795}{317463552}-\frac{426556895}{11337984}\eta \right)\chi^{-19/9}  +              \left(-\frac{14275935425}{416003328}+\frac{209699405}{4000032}\eta \right)\chi^{-25/9}               \right.\nonumber\\
 &\qquad \left.{}    + \left(\frac{198510270125}{10484877312}+\frac{1222893635}{28804608}\eta \right)\chi^{-38/9}  +              \left(\frac{14796093245}{503467776}-\frac{1028884705}{17980992}\eta \right)\chi^{-44/9}          
   \right\rbrace e_0^{4}                                                              \nonumber\\
   &\qquad        +\left\lbrace    \left(-\frac{3 456 734 032 025
}{72 381 689 856
}-\frac{216 909 251 525 }{2 585 060 352
}\eta \right)\chi^{-19/9}  +              \left(-\frac{2 441 897 241 139 735
}{21 246 121 967 616
}+\frac{9 479 155 594 325}{58 368 466 944
}\eta \right)\chi^{-25/9}               \right.\nonumber\\
 &\qquad \left.{}     + \left(\frac{659 649 627 625 375
}{4 781 104 054 272
}+\frac{4 063 675 549 105}{13 134 901 248
}\eta \right)\chi^{-38/9}  +              \left(\frac{1 968 906 345 873 305
}{5 969 113 952 256
}-\frac{8 999 675 405 695}{16 398 664 704
}\eta \right)\chi^{-44/9}                      \right.\nonumber\\     
 &\qquad \left.{}       
       + \left(-\frac{144 936 872 901
}{1 691 582 464
}- \frac{7 378 552 295}{ 32 530 432}\eta \right)\chi^{-19/3}  +    \left(-\frac{213 483 902 125
}{1 117 863 936}+\frac{14 845 156 625}{39 923 712
}\eta \right)\chi^{-7}                   \right\rbrace e_0^{6}    ,           \\  \nonumber\\
 &\mathcal{C}_3=  -16\pi + \left(  \frac{65561 \pi}{4080 }\chi^{-19/9} -\frac{295 945 \pi}{35 088 }\chi^{-28/9}    \right) e_0^2 + \left(\frac{ 217 859 203 \pi}{ 3 720 960}\chi^{-19/9}                   \right.\nonumber\\     
 &\qquad \left.{} 
- \frac{ 3 048 212 305\pi}{ 64 000 512}\chi^{-28/9}- \frac{ 6 211 173 025 \pi}{ 102 085 632}\chi^{-38/9}    +  \frac{ 1 968 982 405 \pi}{ 35 961 984}\chi^{-47/9}    \right)e_0^4     +\left(   \frac{  22 156 798 877 \pi}{ 169 675 776 }\chi^{-19/9}                        \right.\nonumber\\     
 &\qquad \left.{}  -   \frac{  126 468 066 221 755 \pi}{ 846 342 770 688 }\chi^{-28/9}   -   \frac{  20 639 727 962 075 \pi}{46 551 048 192  }\chi^{-38/9}   +    \frac{ 33 366 234 820 475 \pi}{65 594 658 816   }\chi^{-47/9}  \right.\nonumber\\     
 &\qquad \left.{}      +     \frac{ 30 628 811 474 315 \pi}{97 254 162 432   }\chi^{-19/3}    -    \frac{ 28 409 259 125  \pi}{79 847 424  }\chi^{-22/3}     \right)e_0^6  ,       
\end{align}

\begin{align}
&\mathcal{C}_4= \frac{15 293 365}{508 032}+ \frac{27145}{ 504}\eta + \frac{  3085}{72}\eta^2                           + \left\lbrace              \left(- \frac{111064865}{ 14 141 952}-  \frac{165068815}{ 4124736 }\eta - \frac{10 688 155}{294624}\eta^2                            \right)\chi^{-19/9}    +\left( -\frac{5 795 368 945}{350 880 768}      \right.\right.\nonumber\\     
 &\qquad \left.\left.{}                        
    + \frac{4 917 245 }{ 1 566 432}\eta + \frac{25 287 905}{447552}\eta^2                            \right)\chi^{-25/9}     +    \left( \frac{936702035}{1 485 485 568
 }+ \frac{3062285}{ 260064 }\eta - \frac{14 251 675 }{631 584}\eta^2                            \right)\chi^{-31/9}        \right\rbrace   e_0^2    \nonumber\\     
 &\qquad 
+ \left\lbrace  \left(- \frac{ 369068546395}{ 12 897 460 224 }- \frac{ 548523672245}{ 3761759232 }\eta - \frac{35 516 739 065}{  268 697 088
}\eta^2                            \right)\chi^{-19/9}   + \left( -\frac{37 202 269 351 825}{ 297 145 884 672
 }      \right. \right.  \nonumber\\  
  &\qquad \left.  - \frac{2 132 955 527 705}{ 74 286 471 168}\eta+ \frac{ 34 290 527 545 }{ 102 041 856
}\eta^2                            \right)\chi^{-25/9}  
                          + \left(- \frac{94372278903235}{7251965779968 }+ \frac{126823556396665}{ 733 829 870 592
}\eta           
                 \right.  \nonumber\\  
&\qquad\left.   \left.        - \frac{20 940 952 805 }{ 93 768 192}\eta^2       \right)\chi^{-31/9}     +\left( \frac{418677831611033
}{34 573 325 230 080}+ \frac{2163514670909 }{12862100160} \eta + \frac{ 203 366 083 643 }{1 130 734 080}\eta^2                            \right)\chi^{-38/9}                    \right.\nonumber\\    
&\qquad\left. \left.    +  \left(  \frac{562 379 595 264 125
}{5 284 378 165 248}   +  \frac{ 2 965 713 234 395}{94 363 895 808}\eta - \frac{ 240 910 046 095}{518 482 944}\eta^2                            \right)\chi^{-44/9}   +  \left( \frac{3654447011975}{ 98224939008}       \right.\right.\right.\nonumber\\
&\qquad \left.\left.\left.   -  \frac{4300262795285}{ 18 124 839 936
 }\eta  + \frac{ 392 328 884 035 }{1 294 631 424}\eta^2   \right)\chi^{-50/9}  \right\rbrace   e_0^4    \right.       \nonumber\\
&\qquad    +  \left\lbrace     \left(- \frac{187675742904025}{  2 940 620 931 072
 }-  \frac{ 278930807554775}{857681104896}\eta - \frac{ 18 060 683 996 675 }{   61 262 936 064
}\eta^2                     
       \right)\chi^{-19/9}           \right. 
         \nonumber\\
&\qquad              
+ \left(- \frac{ 6 363 444 229 039 638 215}{  15 175 834 621 968 384 }-  \frac{39 088 433 492 776 445}{270 997 046 820 864}\eta + \frac{ 1 550 053 258 427 425 }{1 488 994 762 752}\eta^2                     
       \right)\chi^{-25/9} 
         \nonumber\\
&\qquad              
+ \left(- \frac{387035983120116605285}{5 846 592 827 536 441 344}+ \frac{1095104635088909345 }{1 338 505 683 959 808  }\eta - \frac{ 185 468 261 986 684 025}{ 191 215 097 708 544
}\eta^2                     
       \right)\chi^{-31/9} 
         \nonumber\\
&\qquad              
+ \left( \frac{1391266434443462659}{  15 765 436 304 916 480 }+ \frac{ 7189359251430607}{ 5865117672960 }\eta +   \frac{ 675 785 495 945 689 }{ 515 614 740 480
}\eta^2                     
       \right)\chi^{-38/9} 
         \nonumber\\
&\qquad              
+ \left( \frac{74 835 480 932 061 169 625}{62 651 587 527 180 288
 }+ \frac{ 14 868 442 349 448 515}{21 514 968 244 224
 }\eta - \frac{2 107 245 064 767 505 }{472 856 444 928
}\eta^2                     
       \right)\chi^{-44/9} 
         \nonumber\\
&\qquad              
+ \left( \frac{43949506831840859555}{ 63 177 102 070 677 504 }-  \frac{1344731894414361455 }{376 054 178 992 128
   }\eta + \frac{ 7 946 157 848 161 165}{2 066 231 752 704}\eta^2                     
       \right)\chi^{-50/9} 
       \nonumber\\
&\qquad              
+ \left(- \frac{984783138418096685}{ 40 879 050 017 734 656 }-  \frac{258954290041765}{271268315136 }\eta - \frac{ 173 415 564 792 655  }{148 551 696 384}\eta^2                     
       \right)\chi^{-19/3} 
       \nonumber\\
&\qquad               
+ \left(- \frac{136 868 720 309 511}{ 189 457 235 968 }-  \frac{  17 969 188 685 519}{35 523 231 744  }\eta + \frac{ 1 453 574 802 115 }{ 390 365 184}\eta^2                     
       \right)\chi^{-7} \nonumber\\
&\qquad \left.              
+ \left(- \frac{26945014260125}{52819070976}+ \frac{17350371000625}{6 707 183 616  }\eta - \frac{ 357 715 525 375 }{ 119 771 136}\eta^2                     
       \right)\chi^{-23/3}           \right\rbrace e_0^6   \,. 
\end{align}
\end{subequations}

\newpage

\end{widetext}
The frequency dependence of $e_t$ due to  2PN-accurate GW induced $e_t$ evolution 
is given by

%The 2PN-accurate expression of $e_t$ to be used for the application of SPA is given as follows
\begin{align}                           \label{appendixet}
e_t &\sim 
      \sum_{n=0}^{4} \mathcal{D}_n x^{n/2}\,.    
\end{align}
The coefficients  $\mathcal{D}_n$ that incorporate all the ${\cal O}{\left(e_0^5\right)}$ contributions read
\begin{widetext}

\begin{subequations}
\begin{align}
&\mathcal{D}_0 =   e_0\chi^{-19/18} + \frac{3323}{1824} e_0^3 \left(\chi^{-19/18}- \chi^{-19/6}\right) + 
\left( \frac{15 994 231
}{6 653 952}\chi^{-19/18} -\frac{11 042 329
}{1 108 992} \chi^{-19/6}+\frac{50 259 743
}{6 653 952} \chi^{-95/18}\right)e_0^5 ,  \\    \nonumber\\    
&\mathcal{D}_1 =  0                  ,    \\    \nonumber\\    
&\mathcal{D}_2 =\left\lbrace     \left(\frac{2833}{2016}-\frac{197 }{72}\eta\right)   \left( -\chi^{-19/18}+  \chi^{-31/18}\right)      \right\rbrace e_0  + \left\lbrace   \left(-\frac{9 414 059
}{3 677 184
} +\frac{654 631 }{131 328
 }\eta \right)\chi^{-19/18}        \right.\nonumber\\
 &\qquad \left. +\left(\frac{386 822 573
}{47 803 392
} -\frac{1 482 433 }{131 328
 }\eta\right)\chi^{-31/18}  +\left(\frac{11 412 055}{ 5 311 488} -\frac{ 378 697 }{43 776 }\eta\right)\chi^{-19/6}   \right.\nonumber\\
  &\qquad \left.  +\left(-\frac{9 414 059
}{ 1 225 728} +\frac{ 654631 }{43 776 }\eta\right)\chi^{-23/6}   \right\rbrace e_0^3     
+  \left\lbrace  \left(-\frac{45 311 656 423 }{ 13 414 367 232} +\frac{ 3 150 863 507 }{479 084 544
 }\eta\right)\chi^{-19/18}     \right.     \nonumber\\ 
  &\qquad \left.  +\left(\frac{3 061 519 891 285
}{ 174 386 774 016 } - \frac{  11 147 601 665 }{479 084 544}\eta\right)\chi^{-31/18}  +\left(\frac{ 37 922 258 765}{3 229 384 704
} -\frac{ 1 258 410 131}{26 615 808 }\eta \right)\chi^{-19/6}            \right.\nonumber\\     
  &\qquad \left.     +\left(-\frac{699 589 093 187
}{ 9 688 154 112} +\frac{3 092 267 495 }{26 615 808 }\eta\right)\chi^{-23/6} +  \left(-\frac{1 182 747 028 465
}{ 174 386 774 016 } +\frac{ 24 493 152 461}{479 084 544
 }\eta\right)\chi^{-95/18}    \right.\nonumber\\
    &\qquad \left. +\left(\frac{711 929 259 595
}{ 13 414 367 232
} -\frac{  49 505 846 855 }{ 479 084 544
 }\eta \right)\chi^{-107/18} \right\rbrace e_0^5   ,       \\    \nonumber\\    
&\mathcal{D}_3=   \left\lbrace     \frac{377 \pi}{144}      \left(-\chi^{-19/18}+\chi^{-37/18}  \right) \right\rbrace       e_0 + \left( -\frac{1 252 771 \pi}{ 262 656 }\chi^{-19/18}+\frac{ 1 315 151 \pi}{ 131 328 }\chi^{-37/18} +\frac{396797 \pi}{43776}\chi^{-19/6}            \right.\nonumber\\     
  &\qquad \left.   -   \frac{1 252 771  \pi}{87 552}\chi^{-25/6}     \right)e_0^3  +\left(   -   \frac{6 029 825 087  \pi}{958 169 088}\chi^{-19/18} +  \frac{607 032 981 553   \pi}{27 786 903 552 }\chi^{-37/18}   +  \frac{1 318 556 431   \pi}{26 615 808}\chi^{-19/6}    \right.\nonumber\\
   &\qquad \left.     - \frac{1 422 200 801   \pi}{13 307 904}\chi^{-25/6} -  \frac{1 586 634 546 601   \pi}{27 786 903 552 }\chi^{-95/18}  +  \frac{94 739 615 555  \pi}{958 169 088 }\chi^{-113/18}    \right)e_0^5 ,     \\    \nonumber\\    
&\mathcal{D}_4=    \left\lbrace    \left( \frac{77006005
}{24 385 536}  -\frac{1143767 }{145 152
}\eta + \frac{43 807  }{10368}\eta^2 \right)\chi^{-19/18} + \left( -\frac{8 025 889
}{4 064 256}  +\frac{558 101
}{72 576}\eta  - \frac{ 38 809 }{5184}\eta^2 \right)\chi^{-31/18}      \right.\nonumber\\
  &\qquad \left. + \left( -\frac{28850671}{24 385 536}  +\frac{27565 }{145 152}\eta + \frac{33 811}{10368}\eta^2 \right)\chi^{-43/18}      \right\rbrace e_0
+  \left\lbrace          \left( \frac{255890954615
}{44 479 217 664}  -\frac{3800737741 }{ 264 757 248}\eta     \right.          \right.\nonumber\\
  &\qquad \left.     \left.+ \frac{145 570 661  }{18 911 232}\eta^2 \right)\chi^{-19/18}     + \left( -\frac{1 095 868 349 309
}{96 371 638 272
}  +\frac{65 400 285 919 }{1 720 922 112}\eta - \frac{ 292 039 301  }{9 455 616}\eta^2 \right)\chi^{-31/18}        \right.\nonumber\\
  &\qquad \left.   + \left(- \frac{20952382669619}{4 047 608 807 424}  -\frac{ 385200824731 }{ 24 092 909 568}\eta + \frac{4 301 644 427  }{ 132 378 624}\eta^2 \right)\chi^{-43/18}  + \left( \frac{8180980796033}{ 1 349 202 935 808}    \right.     \right.\nonumber\\
  &\qquad \left.\left.     +\frac{14604819923}{ 2 676 989 952}\eta - \frac{317 361 763  }{ 14 708 736
}\eta^2 \right)\chi^{-19/6}      +   \left( \frac{32 330 351 815
}{3 569 319 936}  -\frac{10 345 778 159 }{191 213 568}\eta + \frac{ 74 603 309  }{ 1 050 624}\eta^2 \right)\chi^{-23/6} \right.\nonumber\\
  &\qquad \left. + \left( -\frac{9164199307}{ 2 118 057 984}  +\frac{1205846917}{29 417 472
}\eta - \frac{13 714 021   }{233 472
}\eta^2 \right)\chi^{-9/2}         \right\rbrace e_0^3              \nonumber\\ 
  &\qquad +  \left\lbrace      \left( \frac{1231651832357155}{162 260 186 038 272} -\frac{18293673608177}{965 834 440 704 }\eta +\frac{700 659 277 417 }{68 988 174 336}\eta^2  \right)\chi^{-19/18}      \right.\nonumber\\
 &\qquad \left. + \left( -\frac{8 673 285 852 010 405
}{351 563 736 416 256} + \frac{ 506 837 220 151 715 }{ 6 277 923 864 576 }\eta - \frac{2 196 077 528 005 }{ 34 494 087 168}\eta^2  \right)\chi^{-31/18}    \right.\nonumber\\
 &\qquad \left. + \left(- \frac{4719697288288984795}{191 953 800 083 275 776} -\frac{4676818769915975}{ 87 890 934 104 064 }\eta +\frac{669 607 180 808 035  }{6 277 923 864 576
}\eta^2  \right)\chi^{-43/18}   \right.\nonumber\\
 &\qquad \left. + \left( \frac{27185399185217659}{820 315 384 971 264
} + \frac{48531816604129}{ 1 627 609 890 816
 }\eta -\frac{1 054 593 138 449 }{8 942 911 488
}\eta^2  \right)\chi^{-19/6}  \right.\nonumber\\
 &\qquad \left. + \left( \frac{2 402 572 738 143 295
}{ 28 211 904 774 144} -\frac{ 55 792 908 667 709 }{ 116 257 849 344 }\eta +\frac{352 402 173 805}{ 638 779 392}\eta^2  \right)\chi^{-23/6}  \right.\nonumber\\
 &\qquad \left. + \left( -\frac{ 25186092424407371}{ 273 438 461 657 088} +\frac{936816311138573 }{ 1 627 609 890 816 }\eta - \frac{1 951 606 822 255 }{2 980 970 496}\eta^2  \right)\chi^{-9/2}  \right.\nonumber\\
 &\qquad \left. + \left(- \frac{7937050519029473999}{191 953 800 083 275 776} -\frac{ 1089957759112387 }{87890934104064 }\eta +\frac{ 1 121 044 759 543 031 }{ 6 277 923 864 576
}\eta^2  \right)\chi^{-95/18}  \right.\nonumber\\
 &\qquad \left. + \left(- \frac{16 753 611 658 206 725
}{351 563 736 416 256
} +\frac{2 837 648 691 484 435  }{6 277 923 864 576
 }\eta - \frac{24 125 755 174 085 }{34 494 087 168
}\eta^2  \right)\chi^{-107/18}  \right.\nonumber\\
 &\qquad \left. + \left( \frac{16 633 441 088 056 655
}{ 162 260 186 038 272} -\frac{79153315354555}{ 137 976 348 672 }\eta +\frac{47 507 268 174 605 }{ 68 988 174 336}\eta^2  \right)\chi^{-119/18}           \right\rbrace e_0^5 .     \\    \nonumber\\\nonumber\\\nonumber\nonumber      
\end{align}
\end{subequations}

\end{widetext}

This expression for $e_t$, as expected, is required while operationalizing 
the $\xi_j$ coefficients and therefore $\tilde h(f)$ and the parameter $x$ in the above two expressions should be evaluated at the stationary point.

 We list below the 2PN order expression for $\phi$ that is required to compute the 
 accumulated number of GW cycles, denoted by `2PN analytic' in Table.~\ref{table:1},
as

\begin{align}               \label{appendixPhi}
\phi \sim   \left( \frac{-1}{32 \eta}  \right)       \sum_{n=-5}^{-1} \mathcal{E}_n x^{n/2},  
\end{align}

where the coefficients $\mathcal{E}_n$  are given by

\begin{widetext}

\begin{subequations}
\begin{align}
 &\mathcal{E}_{-5} = 1- \frac{785}{272}e_0^{2}\chi^{-19/9} + \left( -  \frac{2608555}{248064}\chi^{-19/9}+ \frac{5222765}{386688}\chi^{-38/9} \right)e_0^{4}  + \left(- \frac{1326481225}{56558592}\chi^{-19/9}     \right.\nonumber\\ 
  &\qquad \left.{} + \frac{173355248095}{176329728}\chi^{-38/9}- \frac{226068375}{2957312}\chi^{-19/3}  \right)e_0^{6} ,  \\ \nonumber\\   
 &\mathcal{E}_{-4}= 0     ,  \\ \nonumber\\  
 &\mathcal{E}_{-3} = \frac{3715}{1008} + \frac{55}{12}\eta + \left\lbrace    \left(-\frac{2045665}{225792}-\frac{128365}{8064}\eta \right)\chi^{-19/9}  +              \left(-\frac{2223905}{274176}+\frac{154645}{9792}\eta \right)\chi^{-25/9}   \right\rbrace e_0^{2}               						\nonumber\\
 &\qquad       + \left\lbrace    \left(-\frac{6797744795}{205922304}-\frac{426556895}{7354368}\eta \right)\chi^{-19/9}  +              \left(-\frac{14275935425}{232187904}+\frac{209699405}{2232576}\eta \right)\chi^{-25/9}               \right.\nonumber\\
 &\qquad \left.{}     + \left(\frac{198510270125}{4493518848}+\frac{1222893635}{12344832}\eta \right)\chi^{-38/9}  +              \left(\frac{14796093245}{194890752}-\frac{1028884705}{6960384}\eta \right)\chi^{-44/9}          
   \right\rbrace e_0^{4}                                                              \nonumber\\
   &\qquad       + \left\lbrace    \left(-\frac{3 456 734 032 025
}{46950285312}-\frac{216 909 251 525 }{1676795904
}\eta \right)\chi^{-19/9}  +              \left(-\frac{2 441 897 241 139 735
}{11858300633088
}+\frac{9 479 155 594 325}{32577748992
}\eta \right)\chi^{-25/9}               \right.\nonumber\\
 &\qquad \left.{}     + \left(\frac{659 649 627 625 375
}{2049044594688
}+\frac{4 063 675 549 105}{5629243392
}\eta \right)\chi^{-38/9}  +              \left(\frac{1 968 906 345 873 305
}{2310624755712
}-\frac{8 999 675 405 695}{ 6347870208
}\eta \right)\chi^{-44/9}                      \right.\nonumber\\     
 &\qquad \left.{}       
     +   \left(-\frac{3623421822525
}{13532659712
}- \frac{184463807375}{ 260243456}\eta \right)\chi^{-19/3}  +    \left(-\frac{213 483 902 125
}{331218944}+\frac{14 845 156 625}{11829248
}\eta \right)\chi^{-7}                   \right\rbrace e_0^{6}   ,            \\ \nonumber\\
 &\mathcal{E}_{-2}=  -10\pi + \left(  \frac{65561 \pi}{2880 }\chi^{-19/9} -\frac{295 945 \pi}{19584 }\chi^{-28/9}    \right) e_0^2 + \left(\frac{ 217 859 203 \pi}{2626560}\chi^{-19/9}                   \right.\nonumber\\     
 &\qquad \left.{} 
- \frac{ 3 048 212 305\pi}{ 35721216}\chi^{-28/9}- \frac{ 6 211 173 025 \pi}{ 46227456}\chi^{-38/9}    +  \frac{ 1 968 982 405 \pi}{13920768}\chi^{-47/9}    \right)e_0^4     +\left(   \frac{  22 156 798 877 \pi}{ 119771136 }\chi^{-19/9}                        \right.\nonumber\\     
 &\qquad \left.{}  -   \frac{  126 468 066 221 755 \pi}{ 472377360384 }\chi^{-28/9}   -   \frac{  20 639 727 962 075 \pi}{ 21079719936}\chi^{-38/9}   +    \frac{ 33 366 234 820 475 \pi}{25391480832  }\chi^{-47/9}  \right.\nonumber\\     
 &\qquad \left.{}      +     \frac{ 30 628 811 474 315 \pi}{32418054144}\chi^{-19/3}    -    \frac{ 28 409 259 125  \pi}{23658496}\chi^{-22/3}     \right)e_0^6   ,      \\    \nonumber\\    
&\mathcal{E}_{-1}= \frac{15 293 365}{1016064}+ \frac{27145}{ 1008}\eta + \frac{  3085}{144}\eta^2                           + \left\lbrace              \left(- \frac{111064865}{ 10948608}-  \frac{165068815}{ 3193344}\eta - \frac{10 688 155}{228096}\eta^2                            \right)\chi^{-19/9}                    +\left( -\frac{5 795 368 945}{227598336}      \right.\right.\nonumber\\     
 &\qquad \left.\left.{}                        
    + \frac{4 917 245 }{ 1016064 }\eta + \frac{25 287 905}{290304}\eta^2                            \right)\chi^{-25/9}     +    \left( \frac{936702035}{829108224}+ \frac{3062285}{ 145152}\eta - \frac{14 251 675 }{352512}\eta^2     \right)\chi^{-31/9}        \right\rbrace   e_0^2     \nonumber\\     
 &\qquad +
\left\lbrace  \left(- \frac{ 369068546395}{ 9985130496}- \frac{ 548523672245 }{  2912329728 }\eta - \frac{35 516 739 065}{  208023552
}\eta^2      \right)\chi^{-19/9}   + \left( -\frac{ 37 202 269 351 825}{ 192743276544 }      \right. \right.  \nonumber\\  
  &\qquad \left.  - \frac{2 132 955 527 705}{ 48185819136
}\eta+ \frac{ 34 290 527 545 }{ 66189312
}\eta^2     \right)\chi^{-25/9}  
   + \left(- \frac{94372278903235}{4047608807424 }+ \frac{126823556396665 }{409579462656
}\eta           
                 \right.  \nonumber\\  
&\qquad\left.   \left.        - \frac{900460970615}{ 2250436608}\eta^2       \right)\chi^{-31/9}     +\left( \frac{2093389158055165}{82975980552192}+ \frac{10817573354545 }{30869040384 }\eta + \frac{1016830418215}{2713761792}\eta^2                            \right)\chi^{-38/9}                    \right.\nonumber\\    
&\qquad\left. \left.    +        \left(  \frac{562 379 595 264 125
}{2264733499392}   +  \frac{ 2 965 713 234 395}{40441669632}\eta - \frac{ 240 910 046 095}{222206976}\eta^2                            \right)\chi^{-44/9}   +  \left( \frac{113287857371225}{ 1178699268096
}       \right.\right.\right.\nonumber\\
&\qquad \left.\left.\left.   -  \frac{4300262795285}{ 7016067072}\eta  + \frac{ 392 328 884 035 }{501147648
}\eta^2   \right)\chi^{-50/9}  \right\rbrace   e_0^4    \right.       \nonumber\\
&\qquad    +  \left\lbrace     \left(- \frac{187675742904025}{ 2276609753088
 }-  \frac{  278930807554775 }{664011177984
  }\eta - \frac{ 18 060 683 996 675 }{  47429369856
}\eta^2                     
       \right)\chi^{-19/9}           \right. 
         \nonumber\\
&\qquad              
+ \left(- \frac{ 6 363 444 229 039 638 215}{ 9843784619655168 }-  \frac{39 088 433 492 776 445}{175781868208128}\eta + \frac{ 1 550 053 258 427 425 }{965834440704}\eta^2                     
       \right)\chi^{-25/9} 
         \nonumber\\
&\qquad              
+ \left(- \frac{387035983120116605285}{3263214601415688192}+ \frac{ 1095104635088909345 }{747072939884544 }\eta - \frac{ 185 468 261 986 684 025}{ 106724705697792
}\eta^2                     
       \right)\chi^{-31/9} 
         \nonumber\\
&\qquad              
+ \left( \frac{6956332172217313295}{  37837047131799552 }+ \frac{ 35946796257153035 }{ 14076282415104}\eta + \frac{ 3378927479728445 }{ 1237475377152
}\eta^2                     
       \right)\chi^{-38/9} 
         \nonumber\\
&\qquad              
+ \left( \frac{74 835 480 932 061 169 625}{ 26850680368791552
 }+ \frac{ 14 868 442 349 448 515}{9220700676096
 }\eta - \frac{2 107 245 064 767 505 }{202652762112
}\eta^2                     
       \right)\chi^{-44/9} 
         \nonumber\\
&\qquad              
+ \left( \frac{43949506831840859555}{ 24455652414455808 }-  \frac{ 1344731894414361455}{145569359609856
   }\eta + \frac{ 7 946 157 848 161 165}{ 799831646208}\eta^2                     
       \right)\chi^{-50/9} 
       \nonumber\\
&\qquad              
+ \left(- \frac{ 984783138418096685}{ 14218800006168576 }-  \frac{ 5955948670960595}{2170146521088 }\eta - \frac{ 173 415 564 792 655  }{51670155264
}\eta^2                     
       \right)\chi^{-19/3} 
       \nonumber\\
&\qquad               
+ \left(- \frac{3421718007737775}{ 1515657887744 }-  \frac{  449229717137975 }{ 284185853952  }\eta + \frac{ 36339370052875}{ 3122921472 }\eta^2                     
       \right)\chi^{-7} \nonumber\\
&\qquad \left.              
+ \left(- \frac{26945014260125}{ 15650095104 }+ \frac{17350371000625  }{ 1987313664  }\eta - \frac{ 357 715 525 375 }{ 35487744  }\eta^2                     
       \right)\chi^{-23/3}           \right\rbrace e_0^6  . 
\end{align}
\end{subequations}

\end{widetext}

\section{ Explicit 2PN-accurate expressions for constructing our exact in $e_t$ time-domain approximant}  \label{appendixB}

In this appendix, we list explicitly several 2PN-accurate  expressions that are 
required to implement  temporal evolutions in  
$h_{+}(r,\phi,\dot{r},\dot{\phi}) \big|_{\rm Q}$ and $h_{\times}(r,\phi,\dot{r},\dot{\phi}) \big|_{\rm Q}$.
We begin by displaying 2PN-accurate
parametric
 expressions for incorporating the radial part of the dynamics:

\begin{widetext}
\begin{subequations}
\begin{align}
 \label{Eq:r_dotr_2PN_A}
\frac{c^2\,r} {G\, m}
 &= \frac{1 - e_t\,\cos{u}} {x} \,
	 \biggl\{	1
		+	r^{\rm 1PN}(\eta,\,e_t,\,u)\,x  +	r^{\rm 2PN}(\eta,\,e_t,\,u)\,x^2
\biggr\}\,, \\
\frac{ \dot{r} }{c}
 &= \frac{\sqrt{x} \, e_t \, \sin{u}}{1 - e_t\,\cos{u}} \,
		\biggl \{   	1
			+ 	\dot{r}^{\rm 1PN}(\eta,\,e_t)\,x +	\dot{r}^{\rm 2PN}(\eta,\,e_t,\,u)\,x^2
  \biggr \}\,, 
  \end{align}
  \end{subequations}
where the PN coefficients are given by

\begin{subequations}
\begin{align}
 \label{Eq:r_dotr_2PN_B}
r^{\rm 1PN}(\eta,\,e_t,\,u) &= \frac{-24 + 9\,\eta + \nu\,(18 - 7\,\eta) 
			+ e_t^2\,\bigl[24 - 9\,\eta + \nu\,(-6 + 7\,\eta)\bigr]}{6\,\nu\,(1 - e_t^2)},   \\
 r^{\rm 2PN}(\eta,\,e_t,\,u) &= \frac{1}{72\,\nu\,(1 - e_t^2)^2}\,\biggl\{-288 + 765\,\eta - 27\,\eta^2 
			+ e_t^2\,(288 - 1026\,\eta + 54\,\eta^2) + e_t^4\,(261\,\eta - 27\,\eta^2) \nonumber \\
			&+ \Bigl(-540 + e_t^2\,(540 - 216\,\eta) + 216\,\eta\Bigr)\,\sqrt{1 - e_t^2} 
			+ \nu\,\biggl[648 - 567\,\eta + 35\,\eta^2 \nonumber \\
			&+ e_t^2\,(468 + 150\,\eta - 70\,\eta^2)
		+ e_t^4\,(72 - 231\,\eta + 35\,\eta^2) + \Bigl(180 - 72\,\eta 
		+ e_t^2\,(-180 + 72\,\eta)\Bigr)\,\sqrt{1 - e_t^2}\biggr]\biggr\}\,, \\
\dot{r}^{\rm 1PN}(\eta,\,e_t) &= \frac{-7\,\eta + e_t^2\,(-6 + 7\,\eta)}{6\,(1 - e_t^2)}\,,   \\
 \dot{r}^{\rm 2PN}(\eta,\,e_t,\,u) &= \frac{1}{72\,\nu^3\,(1 - e_t^2)^2}\,\biggl\{
			      -135\,\eta + 9\,\eta^2 + e_t^2\,(405\,\eta - 27\,\eta^2) 
		   + e_t^4\,(-405\,\eta + 27\,\eta^2) + e_t^6\,(135\,\eta - 9\,\eta^2) \nonumber \\
		   &+ \nu\,\Bigl[-540 + 351\,\eta - 9\,\eta^2 + e_t^2\,(1080 - 702\,\eta + 18\,\eta^2)
		   + e_t^4\,(-540 + 351\,\eta - 9\,\eta^2)\Bigr] \nonumber \\
		   &+ \nu^3\,\Bigl[-324 + 189\,\eta + 35\,\eta^2
		   + e_t^2\,(-234 + 366\,\eta - 70\,\eta^2) + e_t^4\,(72 - 231\,\eta + 35\,\eta^2)\Bigr] \nonumber \\
		   &- 36\,\nu^2\,(3 + \nu)\,(1 - e_t^2)\,(-5 + 2\,\eta)\,\sqrt{1 - e_t^2}\biggr\}\,. 
\end{align}		   
\end{subequations}

\end{widetext}
In the above expressions $\nu$ is a shorthand for $ 1 - e_t\,\cos{u}$.

The angular variables $\dot \phi $ and $\phi $ are given by

\begin{subequations}
\begin{align}
 \label{Eq:phi_dotphi_2PN_A}
\dot \phi
 &=	\frac{c^3}{G\,m}\,\frac{\sqrt{1 - e_t^2}\,x^{3/2}}{\left(1 - e_t\,\cos{u}\right)^2}
 %\frac{\sqrt{1 - e_t^2}\,\sqrt{x}}{1 - e_t\,\cos{u}}
	 \,
	 \biggl \{  	1 
		+ 	\dot{\phi}^{\rm 1PN}(\eta,\,e_t,\,u)\,x   \nonumber\\
		&+	\dot{\phi}^{\rm 2PN}(\eta,\,e_t,\,u)\,x^2
\biggr\} \,, \\
%-------------------------------------------------------------------------------------------------------------------
\phi
&= \lambda + W(\eta,\, x,\, e_t,\, u) \,, \\
W &=
	(v - u) + e_t \, \sin u 
	+ W^{\rm 1PN}(\eta,\,e_t,\,u)\,x  \nonumber   \\  
	&+ W^{\rm 2PN}(\eta,\,e_t,\,u)\,x^2\,.
\end{align}
\end{subequations}
The 1PN and 2PN parametric contributions to $\dot \phi$ and $W$  are given by

\begin{widetext}
\begin{subequations}
\begin{align}
 \label{Eq:phi_dotphi_2PN_B}
\dot{\phi}^{\rm 1PN}(\eta,\,e_t,\,u) &= \frac{(-1 + \nu + e_t^2)\,(-4 + \eta)}{\nu\,(1 - e_t^2)}\,,  \\
 \dot{\phi}^{\rm 2PN}(\eta,\,e_t,\,u) &= \frac{1}{12\,\nu^3\,(1 - e_t^2)^2}\,\biggl\{-6\,(1-e_t^2)^3\,(3\,\eta + 2\,\eta^2) 
				  + \nu\,\Bigl[108 + 63\eta + 33\,\eta^2 + e_t^2\,(-216 - 126\eta - 66\,\eta^2) \nonumber \\
				  &+ e_t^4\,(108 + 63\eta + 33\,\eta^2)\Bigr] + \nu^2\,\Bigl[-240 - 31\eta - 29\,\eta^2
				  + e_t^4\,(-48 + 17\eta - 17\,\eta^2) + e_t^2\,(288 + 14\eta + 46\,\eta^2)\Bigr] \nonumber \\
				  &+ \nu^3\,\Bigl[42 + 22\eta + 8\,\eta^2 + e_t^2\,(-147 + 8\eta - 14\,\eta^2)\Bigr]
				  + 18\,\nu^2\,(-2 + \nu + 2\,e_t^2)\,(-5 + 2\eta)\,\sqrt{1 - e_t^2}\biggr\}\,,\\
				  W^{\rm 1PN}(\eta,\,e_t,\,u) &= 3\,\frac{e_t\,\sin{u} + (v - u)_{\rm 1PN}}{1-e_t^2}\,,  \\
  W^{\rm 2PN}(\eta,\,e_t,\,u) &= \frac{e_t\,\sin{u}}{32\,\nu^3\,(1 - e_t^2)^2}\,
			 \biggl\{4\,\nu^2\,\biggl[\nu\,\Bigl(108 + e_t^2\,(102 - 52\,\eta) - 56\,\eta\Bigr)
			- 15\,\eta + \eta^2 + e_t^2\,(30\,\eta - 2\,\eta^2) \nonumber \\
			&+ e_t^4\,(-15\,\eta + \eta^2)\biggr] + \biggl[4\,\eta - 12\,\eta^2
			 + e_t^2\,(-8\,\eta + 24\,\eta^2) + e_t^4\,(4\,\eta - 12\,\eta^2) \nonumber \\
			 &+ \nu\,\Bigl(8 + e_t^2\,(-8 - 144\,\eta) + 144\,\eta\Bigr)
			 + \nu^2\,\Bigl(-8 - 148\,\eta + 12\,\eta^2
					  + e_t^2\,(-\eta + 3\,\eta^2)\Bigr)\biggr]\,\sqrt{1 - e_t^2}\biggr\}\,.
\end{align}
\end{subequations}

\end{widetext}

Clearly, we also need to provide the 2PN-accurate parametric expression for $ v-u$.
As noted in the text, there exists an exact expression for $v-u$ in terms 
of the `angular eccentricity'  $e_{\phi}$  \cite{KG06}:

\begin{align}
\label{Eq:vminu_2PN_A}
v - u &=
	2\, \tan^{-1}
	\left[
	\frac{ \beta_{\phi}\, \sin{u} }{ 1 - \beta_{\phi}\, \cos{u} }
	\right]\,,
\end{align}
where  $\beta_{\phi} = ( 1 - \sqrt{ 1 - e_\phi^2 } ) / e_\phi$.
The 2PN-accurate expression for $ \beta_{\phi}$ can be written as
\begin{align}
\label{Eq:vminu_2PN_B}
\beta_{\phi} &=
	      \frac{1 - \sqrt{1 - e_t^2}}{e_t} + \beta_{\phi}^{\rm 1PN}(\eta,\,e_t)\,x
	      + \beta_{\phi}^{\rm 2PN}(\eta,\,e_t)\,x^2\,,
\end{align}
where the 1PN and 2PN contributions are given by

\begin{widetext}
\begin{subequations}
\begin{align}
 \label{Eq:vminu_2PN_C}
 \beta_{\phi}^{\rm 1PN}(\eta,\,e_t) &= \frac{-4 + \eta +e_t^2\,(8 - 2\,\eta) 
				  + (4 - \eta)\,\sqrt{1 - e_t^2}}{e_t\,\sqrt{1 - e_t^2}}\,,  \\
 \beta_{\phi}^{\rm 2PN}(\eta,\,e_t) &= \frac{1}{96\,e_t\,(1 - e_t^2)^{3/2}}\,\biggl\{
			      -528 - 220\,\eta + 4\,\eta^2 
			      + e_t^2\,(5232 - 1659\,\eta + 177\,\eta^2) \nonumber \\
			      &+ e_t^4\,(-3840 + 2086\,\eta - 178\,\eta^2)
		  + \Bigl[528 + 220\,\eta - 4\,\eta^2
		  + e_t^2\,(288 + 83\,\eta - 41\,\eta^2)\Bigr]\,\sqrt{1 - e_t^2}\biggr\}\,.
\end{align}
\end{subequations}

\end{widetext}
The above listed expressions ensure that the orbital time scale variations in the two 
GW polarization states are treated in a parametric manner.
The temporal evolution of these dynamical variables requires 2PN-accurate Kepler equation
and the GW induced variations in $x$ and $e_t$.

The PN approximation is also employed to derive the differential equations for 
$x$, $e_t$ and $l$. 
We display here (again) these three differential equations as 

\begin{widetext}
\begin{subequations}
\label{Eq:dxetl_A}
\begin{align}
\frac{d  x}{d t}                  
      &= \eta \, \frac{c^3}{G\,m} \, x^5 \,
	\biggl \{	\frac{192 + 584\,e_t^2 + 74\,e_t^4}{15\,(1 - e_t^2)^{7/2}} +	\dot{x}^{\rm 1PN}(\eta,\,e_t)\,x
		+	\dot{x}^{\rm 1.5PN}(e_t)\,x^{3/2}  +	\dot{x}^{\rm 2PN}(\eta,\,e_t)\,x^2
	\biggr\}\,, \\      
\frac{d  e_t}{d t}                    
      &= - \eta \, e_t \, \frac{c^3}{G\,m} \, x^4 \,
	\biggl \{	\frac{304 + 121\,e_t^2}{15\,(1 - e_t^2)^{5/2}} +	\dot{e}_t^{\rm 1PN}(\eta,\,e_t)\,x 
		+	\dot{e}_t^{\rm 1.5PN}(e_t)\,x^{3/2} +	\dot{e}_t^{\rm 2PN}(\eta,\,e_t)\,x^2
	\biggr\}\,, \\              
\frac{d  l}{d t} &= n 
      = x^{3/2} \, \frac{c^3}{G\,m}
	\biggl \{	1
		+	\dot{l}^{\rm 1PN}(e_t)\,x 
		+	\dot{l}^{\rm 2PN}(\eta,\,e_t)\,x^2
	\biggr\}\,, 
\end{align}
\end{subequations}

where the PN coefficients are given by

\begin{subequations}     
\label{Eq:dxetl_B}
\begin{align}
%\label{Eq:1PN2PNcoeff_evoleq}
 \dot{x}^{\rm 1PN}(\eta,\,e_t) &= \frac{-11888 - 14784\,\eta + e_t^2\,(87720 - 159600\,\eta) + e_t^4\,(171038 - 141708\,\eta) 
		    + e_t^6\,(11717 - 8288\,\eta)}{420\,(1 - e_t^2)^{9/2}}\,,  \\
 \dot{x}^{\rm 2PN}(\eta,\,e_t) &= \frac{1}{45360\,(1 - e_t^2)^{11/2}}\,\biggl(
		    -360224 + 4514976\,\eta + 1903104\,\eta^2 
		    + e_t^2\,(-92846560 + 15464736\,\eta + 61282032\,\eta^2) \nonumber \\
		    &+ e_t^4\,(783768 - 207204264\,\eta + 166506060\,\eta^2) 
		    + e_t^6\,(83424402 - 123108426\,\eta + 64828848\,\eta^2) \nonumber \\
		    &+ e_t^8\,(3523113 - 3259980\,\eta + 1964256\,\eta^2) 
		    - 3024\,(96 + 4268\,e_t^2 + 4386\,e_t^4
		    + 175\,e_t^6)\,(-5 + 2\,\eta)\,\sqrt{1 - e_t^2}\biggr)\,, \\
 \dot{e}_t^{\rm 1PN}(\eta,\,e_t) &= - \frac{67608 + 228704\,\eta + e_t^2\,(-718008 + 651252\,\eta) 
			+ e_t^4\,(-125361 + 93184\,\eta)}{2520\,(1 - e_t^2)^{7/2}}\,, \\
				\dot{e}_t^{\rm 2PN}(\eta,\,e_t) &= \frac{1}{30240\,(1 - e_t^2)^{9/2}}\,\biggl(
      -15198032 + 13509360\ \eta + 4548096\,\eta^2
      + e_t^2\,(-36993396 - 35583228\,\eta + 48711348\,\eta^2) \nonumber \\
      &+ e_t^4\,(46579718 - 78112266\,\eta + 42810096\,\eta^2)
      + e_t^6\,(3786543 - 4344852\,\eta + 2758560\,\eta^2) \nonumber \\
      &- 1008\,(2672 + 6963\,e_t^2 + 565\,e_t^4)\,(-5 + 2\,\eta)\,\sqrt{1 - e_t^2}\biggr)\,, \\	
\dot{l}^{\rm 1PN}(e_t) &= - \frac{3}{1 - e_t^2}\,,  \\
 \dot{l}^{\rm 2PN}(\eta,\,e_t) &= \frac{-18 + 28\,\eta + e_t^2\,(-51 + 26\,\eta)}{4\,(1 - e_t^2)^2}\,.
\end{align}
\end{subequations}
%In the above equations,
We note that 
the 1.5PN order tail contributions, namely 
$\dot{x}^{\rm 1.5PN} $
and $\dot{e}_t^{\rm 1.5}$, are provided by Eq.~(\ref{Eq:heredevoleq}).

\end{widetext}

%\bibliography{SHG_2606_AG}

%\addbibresource{mybib.bib}
\bibliography{Amybib}
%    \addbibresource{Amybib.bib} 
%\bibliographystyle{apsrev}

\end{document}